\documentclass[12pt]{article}
\usepackage{xcolor}
\usepackage{textcomp}
\usepackage{my_chet}
\usepackage{amssymb,amsmath,amsbsy,amsthm,mathtools}
\usepackage{braket}
\usepackage[compat=1.0.0]{tikz-feynman}
\usepackage{yhmath}

\usepackage[bb=boondox]{mathalfa}

\newcommand{\cO}{\mathcal{O}}
\newcommand{\bE}{\mathbb{E}}
\newcommand{\bR}{\mathbb{R}}
\newcommand{\be}{\begin{equation}}
\newcommand{\ee}{\end{equation}}
\newcommand{\beq}{\begin{eqnarray}}
\newcommand{\eeq}{\end{eqnarray}}
\usepackage{tikz}

\usepackage[numbers,sort&compress]{natbib}

\usepackage{flip-acronyms} \usepackage{tikzfeynman,contour}

\newcommand{\bP}{\mathbb{P}}

\newcommand{\bk}[1]{\langle #1 \rangle}

\newcommand{\ibp}{$\mathcal{IBP}$\,}

\usepackage{listings}
\usepackage{color}

\newcommand{\NC}{\text{NC}}
\newcommand{\PNC}{{\bP}\NC}

\definecolor{dkgreen}{rgb}{0,0.6,0}
\definecolor{gray}{rgb}{0.5,0.5,0.5}
\definecolor{mauve}{rgb}{0.58,0,0.82}

\title{Conformal Fields from Neural Networks}
\author{James Halverson$^{2,3}$\email{j.halverson@northeastern.edu}, Joydeep Naskar$^{2,3}$\email{naskar.j@northeastern.edu}, and Jiahua Tian$^{1}$\email{jhtian@phy.ecnu.edu.cn}}

\affiliation{$^{1}$School of Physics and Electronic Science, East China Normal University,\\ Shanghai 200241, China\\ \vspace{.3cm}$^{2}$Department of Physics, Northeastern University,\\ Boston, MA 02115 USA\\ \vspace{.3cm}$^{3}$The NSF AI Institute for Artificial Intelligence and Fundamental Interactions,\\ Cambridge, MA,
U.S.A}

\abstract{We use the embedding formalism to construct conformal fields in $D$ dimensions, by restricting Lorentz-invariant ensembles of homogeneous neural networks in $(D+2)$ dimensions to the projective null cone. Conformal correlators may be computed using the parameter space description of the neural network. Exact four-point correlators are computed in a number of examples, and we perform a 4D conformal block decomposition that elucidates the spectrum. In a non-unitary example the decomposition precisely matches OPE coefficients for the self-correlator, but not for the mixed correlator. In others, the analysis is facilitated by recent approaches to Feynman integrals. Generalized free CFTs are constructed using the infinite-width Gaussian process limit of the neural network, enabling a realization of the free boson. The extension to deep networks constructs conformal fields at each subsequent layer, with recursion relations relating their conformal dimensions and four-point functions. Numerical approaches are discussed.
}

\begin{document}
\maketitle

\tableofcontents\newpage

\section{Introduction}

Conformal field theories (CFTs) are of central importance in theoretical physics. They describe phase transitions in many systems, providing an organizing principle through the notion of universality, characterized by critical exponents. For instance, the Wilson-Fisher fixed point \cite{PhysRevLett.28.240} describes numerous physical systems (e.g., \cite{PhysRevB.21.3976}), including the liquid-gas phase transition and superfluidity in helium. Two dimensional CFTs enjoy enhanced symmetries  \cite{BELAVIN1984333} and play a prominent role in string theory~\cite{Polchinski:1998rq}, but recent years have seen a renewed interest in higher-dimensional CFTs through the numerical bootstrap program \cite{Rattazzi:2008pe}; see \cite{DiFrancesco:1997nk} and \cite{Fradkin:1996is,Fradkin:1997df} for extensive reviews of 2D and higher-$D$ CFTs (see \cite{Poland:2018epd, Rychkov:2023wsd} for modern developments), respectively. More formally, CFTs provide a framework for understanding field theories, as points on a renormalization group flows between CFT fixed points in the ultraviolet and infrared.

 We provide a new construction of conformal fields \footnote{See the end of the Introduction for clarification on this point.} on $\bR^D$ that utilizes the embedding formalism. This formalism, originally due to Dirac \cite{dirac1936}, notes that $SO(D+1,1)$ is the global conformal group on $\bR^D$ and also the Lorentz group on $\bR^{D+1,1}$. Canonically, one restricts from the Minkowski space $\bR^{D+1,1}$ to the Euclidean subspace $\bR^D$ by first passing to the null cone 
\begin{equation}
\NC := \{ x \cdot x + X_{d+1}^2 - X_0^2 = x \cdot x - X_+ X_- = 0\},
\end{equation}
where $X_\mu = (X_0, x, X_{d+1})\in \bR^{D+1,1}$, the lightcone coordinates are $X_\pm = X_0 \pm X_{d+1}$, and $x \in \bR^D$. The projective null cone is defined to be
\begin{equation}
\PNC = \frac{\NC \setminus 0}{\bR\setminus 0},
\end{equation}
with $X\in \NC$ promoted to homogeneous coordinates identified as $X \sim \lambda X$. The projective scaling maybe be used to set either $X_+=1$ or $X_-=1$, yielding two copies of $\bR^D$ that intersect along a sphere $S^{D-1}$. The $X_+=1$ patch is known as the Poincar\'e section, and on this patch we have 
\begin{equation}
X_\mu = (X_+, x, X_-) = (1, x, x^2) \in \bR^D \subsetneq \bR^{D+1,1}.
\end{equation}
Lorentz boosts violate the condition $X_+=1$, requiring a projective rescaling to arrive back in the Poincar\'e section $\bR^D$, yielding a non-linear realization of the boosts, and therefore the Lorentz group, on $\bR^D$. These are also global conformal transformations on $\bR^D$, a fact much exploited for the study of conformal kinematics; see \cite{Cornalba:2009ax,Weinberg:2010fx,Costa:2011mg} and the lectures \cite{Rychkov:2016iqz}.

Our construction instead utilizes the embedding formalism to \emph{construct} conformal fields, relying crucially on a neural network approach to field theory.
We begin with a Lorentz-invariant theory defined by 
\begin{equation}
Z[J] = \bk{e^{\int d^{D+2}X \, J(X) \Phi(X)}},
\end{equation}
where we leave the expectation general for the moment. The theory may be put on the projective null cone by further specifying $J(X) \mapsto J(X) \delta(X^2)$, yielding
\begin{equation}
    Z[J] = \bk{e^{\int d^{D+2}X \, J(X) \delta(X^2)\, \Phi(X)}},
\end{equation}
together with taking the partition function to be invariant under projective rescaling,
$Z[J(X)] = Z[J(\lambda X)]$,  requiring 
that the integral 
\begin{equation}
    I = \int d^{D+2}X \,\, J(X)\delta(X^2)\, \Phi(X)
\end{equation}
is scale-invariant. This occurs when $J$ and $\Phi$ are homogeneous of degrees $\Delta_J$ and $-\Delta_\Phi$
satisfying
\begin{equation}
D + \Delta_J - \Delta_\Phi = 0,
\end{equation}
where we have used a $-$ sign in the definition so that $\Delta_\Phi$ will eventually be a conformal scaling dimension.
On the null-cone, we may use the fact that $X_\pm = x^2/ X_\mp$ to write the homogeneous field $\Phi(X)$ in terms of one light-cone coordinate $X_\pm$ and $x$, 
either of which yields a field $\phi(x)$ on $\bR^D$ after restricting to the relevant section $X_\pm = 1$. $Z[J]$ endows $\phi(x)$ with associated correlators, yielding a conformally invariant field on $\bR^D$.

The essential step in the construction is to define a Lorentz-invariant field theory of \emph{homogeneous} fields on $\bR^{D+1,1}$. Restricting to homogeneous fields, or any other requirement on the functional form, is not something that we normally do in field theory. For instance, if we define the $\bk{\cdot}$ in $Z[J]$ to be the Feynman path integral, the action $S[\Phi]$ defines a density on fields, but does not explicitly restrict their form. On the other hand, if one were to specify the functional form for $\Phi(X)$, there would be a question of how to endow it with statistics, to give meaning to the $\bk{\cdot}$ in $Z[J]$. This, fortunately, has a solution. Let $\Phi_\Theta(X)$ be a family of functions (with parameters $\Theta$) of fixed functional form. Choosing a parameter density\footnote{Or analytic continuation thereof. We will see that such subtleties are essential to the construction, and that $P(\Theta)$ need not be a probability density, though it is in conventional neural network applications.}  $P(\Theta)$ endows the family with statistics, and we have a partition function
\begin{equation}\label{eq:general_partition_function}
Z[J] = \int D\Theta \, P(\Theta)\, e^{\int d^{D+2} X\, J(X) \Phi_\Theta(X)}
\end{equation}
that defines a field theory. This type of field theory is known as a neural network field theory, and the choice of functional form for $\Phi_\Theta$ is known as the choice of architecture; henceforth, we omit the $\Theta$ subscript, leaving the parameter dependence implicit. To achieve a homogeneous field in this approach, we simply choose a homogeneous architecture.

We view our approach as complementary to recent efforts to study CFTs with the numerical bootstrap \cite{Rattazzi:2008pe,PhysRevD.86.025022, Rattazzi_2011, PhysRevLett.112.141601, Kos:2013tga, Fitzpatrick:2014vua, Fitzpatrick:2012yx, Hogervorst:2013sma, Kos:2016ysd,Beem:2013qxa,El-Showk:2014dwa, Kos:2014bka, Nakayama:2014yia, Nakayama:2016jhq,Poland:2011ey,Chang:2024whx,Caron-Huot:2021rmr,Chester:2019ifh} or applied machine learning \cite{Kantor:2021jpz,Kantor:2021kbx,Kantor:2022epi}, or more abstract mathematical ideas of hyperbolic orbifolds \cite{Kravchuk:2021akc,Bonifacio:2023ban, Gesteau:2023brw}. Whereas some of those techniques using strong constraints from conformal symmetry and unitarity to bound CFT data, we instead use the embedding formalism to construct specific conformal fields directly. Our construction seems to have a significant amount of flexibility, perhaps due to the fact that we do not yet know how to impose unitarity in the form of reflection positivity. Putting these different approaches together is an interesting possibility for future work, and we will discuss it at more length in the Discussion and Outlook. Readers unfamiliar with connections between neural networks and field theory might consult \cite{Halverson:2024hax} for an ML perspective and \cite{Demirtas:2023fir} for a physics perspective, the Introduction of which has a thorough discussion of some of the original literature, e.g., \cite{Halverson_2021,maiti2021symmetryviaduality,halverson2021building}.

\emph{What Constitutes a CFT?} Finally, we would like to clarify aspects of our construction. Our construction of conformal fields provides the sine qua non of a CFT, an ensemble of fields with conformally invariant correlation functions. We take this as our definition: a CFT is a field theory with conformally invariant correlators. Of course, in some circumstances, especially as studied in the bootstrap community, a CFT may have additional structure, e.g. a stress tensor, unitarity, and an operator product expansion. Famous examples exist of CFTs that do not have a stress tensor and / or are non-unitary, but we do not know of an example that does not have an OPE. Conversely, we do not know of a rigorous proof that \emph{any} field theory with conformally invariant correlators necessarily has a conventional OPE, though of course under additional assumptions it may be shown. This is an interesting question, prompted by useful comments from a referee, that we think is a question for the broader community. Nevertheless, despite not knowing the answer in general, we were able to determine the OPE in our non-unitary example, and found a precise match between the OPE coefficients and $D=4$ conformal block decomposition of the self-correlator. However, matching the mixed correlator requires an additional constraint on the moments of $P(\Theta)$ that are not required by conformal invariance alone.

This paper is organized as follows.
In Section \ref{sec:essential_cft} we review the essentials of the embedding formalism and CFT techniques. Our notation conventions are in Appendix \ref{app:notation}.
In Section \ref{sec:NNCFT} we introduce our construction, which relies crucially on three properties: homogeneity, Lorentz invariance in $(D+2)$-dimensions, and finiteness of correlators. We also exactly solve a simple non-unitary theory, demonstrate how free theories may be obtained at large-$N$, and explain how deep neural networks can lead to recursive conformal fields.
In Section \ref{sec:other_treatments} we discuss other treatments of the Lorentzian theory that allow for the study of associated conformal fields. Amplitudes techniques including IBP identities and associated differential equations are used to study a theory satisfying the unitarity bound. Potential numerical approaches are discussed.
In Section \ref{sec:freeboson} we use the techniques we developed to study the free-boson via a large-$N$ limit. We discuss ways in which the free boson can be ``mixed" into interacting theories to ensure certain properties of the conformal block expansion.
In Section \ref{sec:discussion} we discuss our construction further and provide an outlook for future work.

\section{Essential CFT Techniques\label{sec:essential_cft}}

\subsection{The Embedding Formalism}

We review the embedding formalism, elaborating on the discussion and notation presented in the Introduction. See \cite{Rychkov:2016iqz,Poland:2018epd,Costa:2011mg} for relevant references in the Physics literature, and \cite{Schottenloher:1997pw} for a precise mathematically-oriented textbook.

The essential aspect we focus on is how $(D+2)$ dimensional Lorentz transformations can induce conformal transformations on the Poincar\'e section $\bR^D$ given by $X_+=1$. In Minkowski coordinates, we have 
\begin{equation}
    X_\mu = \left(\frac{1+x^2}{2},x,\frac{1-x^2}{2}\right)
\label{eqn:X_ps}
\end{equation}
on the Poincar\'e section, which is $X_\mu = (1,x,x^2)$ in light-cone coordinates. We will act on $X$ with Lorentz transformations and identify the associated conformal transformations.

We begin with the rotation subgroup $SO(D)$ of the conformal group, which arises trivially. Take the Lorentz transformation
\begin{equation}
\Lambda_R = 
\begin{pmatrix}
1 & 0 & 0 \\
0 & R & 0 \\
0 & 0 & 1 \\
\end{pmatrix}
\end{equation}
where $R\in SO(D)$. This clearly induces $x\to R x$ in \eqref{eqn:X_ps}, a rotation in $\bR^D$.

Translations in $\bR^D$ arise via Lorentz transformations of the form
\begin{equation}
\Lambda_T = 
\begin{pmatrix}
1 + \frac{a^2}{2} & a & \cdots & 0 & \frac{a^2}{2} \\
a & 1 & \cdots & 0 & a \\
\vdots & \vdots & \ddots & \vdots & \vdots \\
0 & 0 & \cdots & 1 & 0 \\
-\frac{a^2}{2} & -a & \cdots & 0 & 1 - \frac{a^2}{2}X
\end{pmatrix}.
\end{equation}
Its action on \eqref{eqn:X_ps} is given by
\begin{equation}
\Lambda_{T} X = \left(\frac{1 + (x+a \,e_1)^2}{2},x+a\, e_1,\frac{1 - (x+a \,e_1)^2}{2}\right),
\end{equation}
which is still on the Poincar\'e section, and we see it has induced a translation of $x$ by $a e_1$, where $e_1$ is the unit vector in the $x_1$ direction. By appropriate permutations in $\Lambda_T$ one may induce translations in any of the $D$ directions in $\bR^D$.

\begin{table}[t]
\centering
\begin{tabular}{|c|c|}
\hline
\textbf{Lorentz Generator} & \textbf{Conformal Transformation} \\
\hline
$L_{ij}$ & Rotation \\
$L_{+-}$ & Scaling \\
$L_{i+}$ & Translation \\
$L_{i-}$ & Special Conformal \\
\hline
\end{tabular}
\caption{Lorentz generators in $(D+2)$-dimensions and the conformal transformation they induce on the Poincar\'e section $\bR^D$. In the table $L_{ij}$, $L_{+-}$, $L_{i+}$ and $L_{i-}$ are $\Lambda_R$, $\Lambda_D$, $\Lambda_T$ and $\Lambda_S$, respectively, written in the light-cone coordinates instead of Minkowski, where $i = 1,\cdots,D$ and $+/-$ are the light-cone indices.}
\label{fig:lor_conf}
\end{table}

Special conformal transformations in $\bR^D$ arise via Lorentz transformations
\begin{equation}
\Lambda_S = 
\begin{pmatrix}
1 + \frac{b^2}{2} & b & \cdots & 0 & -\frac{b^2}{2} \\
b & 1 & \cdots & 0 & -b \\
\vdots & \vdots & \ddots & \vdots & \vdots \\
0 & 0 & \cdots & 1 & 0 \\
\frac{b^2}{2} & b & \cdots & 0 & 1 - \frac{b^2}{2}
\end{pmatrix}.
\end{equation}
Its action on \eqref{eqn:X_ps} is given by
\begin{equation}
\Lambda_{S} X = \left(\frac{K + x^2}{2},x+bx^2\, e_1,\frac{K - x^2}{2}\right),
\end{equation}
where $K=1+2bx_1 + b^2 x^2$. In these expressions $x\in \bR^D$, $x_1$ is its first component, and $x^2$ is its length-squared. We see that $(\Lambda_S X)_+=K$ and therefore $\Lambda_S X$ is not on the Poincar\'e section. Using the freedom to projectively rescale by $\lambda = 1/K$, we obtain
\begin{equation}
\frac{1}{K}\Lambda_{S} X = \left(\frac{1 + x^2/K}{2},\frac{x+bx^2\, e_1}{K},\frac{1 - x^2/K}{2}\right),
\end{equation}
which is back on the Poincar\'e section. We see the effective action of $\frac{1}{K}\Lambda_{S} X$ on $x$ is 
\begin{equation}
x \mapsto\frac{x+bx^2\, e_1}{1+2bx_1 + b^2 x^2},
\end{equation}
which is a special conformal transformation associated to the $x_1$ direction. By appropriate permutation in $\Lambda_S$, one can obtain special conformal transformations associated to any of the $D$ directions in $\bR^D$.

Dilatations in $\bR^D$ arise via Lorentz transformations of the form
\begin{equation}
\Lambda_D = 
\begin{pmatrix}
\frac{1+r^2}{2r} & 0 & \cdots & 0 & \frac{1-r^2}{2r} \\
0 & 1 & \cdots & 0 & 0 \\
\vdots & \vdots & \ddots & \vdots & \vdots \\
0 & 0 & \cdots & 1 & 0 \\
\frac{1-r^2}{2r} & 0 & \cdots & 0 & \frac{1+r^2}{2r}
\end{pmatrix}
\end{equation}
which act on $X$ in \eqref{eqn:X_ps} to give $\Lambda_D X = (\frac{1+r^2x^2}{2r},x,\frac{1-r^2x^2}{2r})$. This has $(\Lambda_DX)_+=1/r$ and is no longer in the Poincar\'e section. To arrive back in it, we use a projective scaling by $\lambda=r$ to obtain
\begin{equation}
r\Lambda_D X  = \left(\frac{1+r^2x^2}{2},rx,\frac{1-r^2x^2}{2}\right)
\end{equation}
from which we see that $\Lambda_D$ induces a dilatation on the Poincar\'e section.

In summary, the relationship between Lorentz transformations and non-linearly realized conformal transformations on the Poincar\'e section is given in Table~\ref{fig:lor_conf}.

\subsection{Correlation Functions, Conformal Blocks, and the Stress Tensor}
The two and three-point functions of a conformal field theory are fixed by the conformal symmetry. The two-point function of two scalar operators $\phi_1$ and $\phi_2$ of scaling dimensions $\Delta_1$ and $\Delta_2$ respectively, are given by 
\begin{equation}\label{eq:2pt-scalar}
    \langle \phi_1(x_1) \phi_2(x_2) \rangle = \begin{cases} 
   \frac{1}{(x_{12})^{2\Delta_1}}=\frac{1}{(X_1 \cdot X_2)^{\Delta_1}} & \text{if $\Delta_1=\Delta_2$}, \\
    0 & \text{if $\Delta_1\neq\Delta_2$}.
\end{cases}
\end{equation}
where $X_1\cdot X_2 = -\frac12 (x_1-x_2)^2=: -\frac12 x_{12}^2$, and in the second equality, we suppressed the numerical factor in going from $X^{D+2}$ to $X^{D}$ for simplicity.  The three-point function of three scalar operators $\phi_1$, $\phi_2$ and $\phi_3$ of scaling dimensions $\Delta_1$, $\Delta_2$ and $\Delta_3$ respectively is given by
\begin{equation}\label{eq:3pt-scalar}
\begin{aligned}
    \langle \phi_1(x_1) \phi_2(x_2) \phi_3(x_3) \rangle & = \frac{\lambda_{123}}{(x_{12})^{\Delta_1+\Delta_2-\Delta_3} (x_{13})^{\Delta_1+\Delta_3-\Delta_2} (x_{23})^{\Delta_2+\Delta_3-\Delta_1}} \\
    &=\frac{\lambda_{123}}{(X_1\cdot X_2)^{\frac{\Delta_1+\Delta_2-\Delta_3}{2}} (X_1\cdot X_3)^{\frac{\Delta_1+\Delta_3-\Delta_2}{2}} (X_2 \cdot X_3)^{\frac{\Delta_2+\Delta_3-\Delta_1}{2}}},
\end{aligned}
\end{equation}
where the constant $\lambda_{123}$ is the OPE coefficient. The higher-point functions, in general may involve arbitrary functions of cross-ratios
\begin{equation}
    u=\frac{x_{12}^2 x_{34}^2}{x_{13}^2 x_{24}^2}=\frac{(X_1\cdot X_2) (X_3 \cdot X_4)}{(X_1\cdot X_3) (X_2 \cdot X_4)}, \quad v=\frac{x_{14}^2 x_{23}^2}{x_{13}^2 x_{24}^2}=\frac{(X_1\cdot X_4) (X_2 \cdot X_3)}{(X_1\cdot X_3) (X_2 \cdot X_4)},
\end{equation}
a definition we will use throughout the text.
The four-point function of four scalar operators $\langle \phi_1(x_1) \phi_2(x_2) \phi_3(x_3) \phi_4(x_4)\rangle$ with scaling dimensions $\Delta_1$, $\Delta_2$, $\Delta_3$ and $\Delta_4$ respectively is given by 
\begin{equation}\label{eq:4pt-general}
    \langle \phi_1(x_1) \phi_2(x_2) \phi_3(x_3) \phi_4(x_4)\rangle= \left(\frac{|x_{24}|}{|x_{14}|}\right)^{\Delta_1-\Delta_2} \left(\frac{|x_{14}|}{|x_{13}|}\right)^{\Delta_3-\Delta_4}\frac{g(u,v)}{|x_{12}|^{\Delta_1+\Delta_2}|x_{34}|^{\Delta_3+\Delta_4}}
\end{equation}
where $g(u,v)$ is a function of the cross-ratios.

The function $g(u,v)$ can be decomposed as a sum of conformal blocks of primaries $\mathcal{O}$ that appear both in the $\phi_1\times\phi_2$ and $\phi_3\times\phi_4$ OPEs
\begin{equation}
g(u,v)=\sum_{\mathcal{O}}\lambda_{12\mathcal{O}}\lambda_{34\mathcal{O}}g_{\mathcal{O}}(u,v)
\end{equation}
where $g_{\mathcal{O}}(u,v)$ is the conformal block associated with primary $g_{\mathcal{O}}$.
\newline

Let $\phi$ be the lowest scalar primary of scaling dimension $\Delta$ in a $D$-dimensional unitary theory. By unitarity bound, we have $\Delta\geq\frac{D-2}{2}$. 
Let us consider the four point function $\langle \phi(x_1)\phi(x_2)\phi(x_3)\phi(x_4)\rangle$ and express it in terms of the cross-ratios
\begin{equation}\label{eq:4ptallphi}
        \langle \phi(x_1)\phi(x_2)\phi(x_3)\phi(x_4)\rangle=\frac{g(u,v)}{x_{12}^{2\Delta} x_{34}^{2\Delta}}
\end{equation}
where $g(u,v)$ satisfies 
\begin{align}
    \label{eq:crossing}
    g(u,v) = \left(\frac{u}{v}\right)^{\Delta_\Phi}g(v,u) = g(u/v,1/v) 
\end{align}
which follow from crossing symmetry.
The conformal block decomposition (CBD) takes the form
\begin{equation}\label{eq:allphig}
g(u,v)= \sum_{\mathcal{O}\in\phi\times\phi}\lambda_{\phi\phi\mathcal{O}}^2 g_{\mathcal{O}}(u,v)
\end{equation}
and may be used to better understand the operator content of the theory.

Suppose that this unitary theory has a well-defined local energy-momentum tensor. In that case, \ref{eq:allphig} contains the energy-momentum tensor operator $\mathcal{O}_{D,2}$ having a spin $2$ and scaling dimension $D$. The recipe to extract each $\lambda^2_{\phi\phi \mathcal{O}_{2,D}}$ is to use equations~(\ref{eq:4ptallphi}) and~(\ref{eq:allphig}) and write
\begin{equation}
    \langle \phi(x_1)\phi(x_2)\phi(x_3)\phi(x_4)\rangle {x_{12}^{2\Delta} x_{34}^{2\Delta}} = g(u,v)= \sum_{\mathcal{O}\in\phi\times\phi}\lambda_{\phi\phi\mathcal{O}}^2 g_{\mathcal{O}}(u,v)
\end{equation}
and read-off the coefficients $\lambda_{\phi\phi\mathcal{O}}^2$. Naively, it appears we need an explicit form of $g(u,v)$, the expression for conformal blocks $g_{\mathcal{O}}$, and the operator spectrum of the $\phi\times\phi$ OPE. Fortunately, we do not always need the information about operator spectrum. One can find the operator spectrum as well as the OPE coefficients by an appropriate expansion \cite{Dolan:2000ut}. Reviewing the essentials, it is useful to change variables to
\begin{equation}\label{eq:zx-variables}
    u=z\Bar{z}, \quad v=(1-z)(1-\Bar{z}),
\end{equation}
or, equivalently $z,\Bar{z}$ are defined as
\begin{equation}\label{eq:z-zbar-def}
    z,\Bar{z}=\frac{1}{2}(u-v+1\pm \sqrt{(u-v+1)^2-4u}).
\end{equation}
The recipe is to Taylor expand the function $g(z,\Bar{z})$ near $z=0_+, \Bar{z}=0_+$\footnote{For $D=2,4$ we can expand separately in $z$ and $\Bar{z}$, but for odd $D$, we are limited in our abilities to expand around $z=\Bar{z}$ for not having a closed form expression for the conformal blocks.}. One then expands the conformal blocks for \emph{various} primaries\footnote{The leading order power of the conformal block expansion depends on the value of $\Delta$ and $l$. So only a discrete set of operators with specific values of $\Delta$ and $l$ can contribute to this sum. The expansion should be carried out over the conformal blocks of those operators having allowed values of $\Delta$ and $l$. Moreover, if the constraint equation at a given order is satisfied with lower $\Delta$ operators, higher $\Delta$ operators do not contribute.} and match the expansion on LHS and RHS at every order of $x$ and $z$. 
\begin{equation}
    \sum_{m,n=0}^{\infty} g_{mn} z^m \Bar{z}^n = \sum_{\mathcal{O}\in\phi\times\phi} \lambda^2_{\phi\phi\mathcal{O}} \sum_{m,n=0}^{\infty} g_{{\mathcal{O}}{mn}} z^m \Bar{z}^n
\end{equation}
where $g_{mn}$ and $g_{\mathcal{O}mn}$ denote the coefficients of $g(z,\Bar{z})$ and $g_{\mathcal{O}}(z,\Bar{z})$ at order $z^m \Bar{z}^n$ respectively. For a fixed pair of $m$ and $n$, we have
\begin{equation}
    g_{mn}= \sum_{\mathcal{O}\in\phi\times\phi} \lambda^2_{\phi\phi\mathcal{O}} g_{{\mathcal{O}}{mn}}
\end{equation}
One can solve for $\lambda^2_{\phi\phi\mathcal{O}}$ recursively. The energy-momentum tensor $\mathcal{O}_{2,D}$ contributes to the RHS expansion and one can extract the coefficient $\lambda^2_{\phi\phi O_{2,D}}$. Using this, we can calculate the value of the central charge $c$. 

As an example, we illustrate this with the example of free scalar theory in four dimensions. In this case we have
\begin{equation}
    g(u,v)=1+u+\frac{u}{v}=1+z\Bar{z}+\frac{z\Bar{z}}{(1-z)(1-\Bar{z})}
\end{equation}
The general expression for the conformal block in four dimensions for an operator $\mathcal{O}$ of spin $l$ and scaling dimension $\Delta$ is given by\footnote{We follow the normalization used in \cite{Rattazzi:2008pe}.}
\begin{equation}
    g_{\Delta,l}(z,\Bar{z})=\left(\frac{-1}{2}\right)^l \frac{z\Bar{z}}{\Bar{z}-z}\left( k_{\Delta+l}(\Bar{z})k_{\Delta-l-2}(z) - \Bar{z} \leftrightarrow z \right)
\end{equation}
where:
\begin{equation}
    k_{\beta}(x)=x^{\beta/2}\, _2F_1 \left(\frac{\beta}{2}, \frac{\beta}{2}; \beta, x \right)
\end{equation}
We note that more generally for the CBD of $\langle 
\cO_1(x_1) \cO_2(x_2) \cO_3(x_3) \cO_4(x_4) \rangle$, we use the following basic function~\cite{Dolan:2003hv}:
\begin{equation}\label{eq:basic_function}
    k_{\beta}(x)=x^{\beta/2}\, _2F_1 \left(\frac{\beta - \Delta_{12}}{2}, \frac{\beta + \Delta_{34}}{2}; \beta, x \right)
\end{equation}
where $\Delta_{ij} \equiv \Delta_i - \Delta_j$ for $\Delta_i$ the conformal dimension of $\cO_i$. It is easy to check that in the 4d free theory that only operators with $\Delta-l=2$ contribute to the $\phi\times\phi$ OPE with OPE coefficients given by 
\begin{equation}
    \lambda^2_{\phi\phi\mathcal{O}}(l=2n, \Delta=l+2)\equiv p_{l,l+2} = 2^{l+1} \frac{(l!)^2}{(2l)!}
\end{equation}
We find that $\lambda^2_{\phi\phi\mathcal{O}_{2,4}}=4/3$. The central charge is thus found to be
\begin{equation}
    c=\frac{16}{9}\frac{\Delta^2}{\lambda^2_{\phi\phi\mathcal{O}_{2,4}}}=\frac{4}{3}
\end{equation}
which is indeed the correct value of central charge for free scalar theory in four dimensions.

\section{Conformal Fields from Neural Networks \label{sec:NNCFT}}

Let us recall the outline of our construction presented in the Introduction and elaborate. 

The starting point is a Lorentz-invariant field theory
\begin{equation}
Z[J] = \int D\Theta \, P(\Theta)\, e^{\int d^{D+2} X\, J(X) \Phi(X)}
\end{equation}
where $\Phi$ depends on $\Theta$. In concrete examples, Lorentz invariance of $Z[J]$ may be demonstrated by absorbing a Lorentz transformation of $X$ into a redefinition of some or all of the parameters and using Lorentz invariance of $P(\Theta)$. Correlators may be computed in parameter space in the usual way by appropriate $J$-derivatives, yielding
\begin{equation}
G^{(n)}(X_1,\dots,X_n) = \int D\Theta \, P(\Theta)\, \Phi(X_1)\dots\Phi(X_n).
\end{equation}
We henceforth use 
\begin{equation}
\bk{\mathcal{O}} = \int D\Theta \, P(\Theta)\, \mathcal{O}
\end{equation}
to denote parameter space expectations. Sometimes $P(\Theta)$ is a proper probability distribution. More broadly it is a function to integrate $\mathcal{O}$ against to define the expectation.

We obtain a conformal field by restriction to the Poincar\'e section (PS) $\bR^D$ if $\Phi$ is homogeneous
\begin{equation}
\Phi(\lambda X) = \lambda^{-\Delta_\Phi} \, \Phi(X).
\end{equation}
This condition ensures well-definedness of the partition function on the PNC and also that of the conformal field
\begin{equation}
\phi(x) := \Phi(X)|_\text{PS}
\end{equation}
on the Poincar\'e section transforms appropriately under scale transformations. The Euclidean conformal correlators on $\bR^D$ are simply the restriction of the Lorentzian correlators on $\bR^{D+1,1}$
\begin{equation}\label{eq:Gx_from_GX}
\begin{split}
    G^{(n)}(X_1, \ldots, X_n) &= \bk{\Phi(X_1)\dots\Phi(X_n)}, \\ 
    G^{(n)}(x_1, \ldots, x_n) &:= G^{(n)}(X_1|_{PS}, \ldots, X_n|_{PS}) = \bk{\phi(x_1) \cdots \phi(x_n)}.    
\end{split}
\end{equation}
If one additionally requires $\phi$ to be a conformal primary, then an additional scaling relation corresponding to special conformal transformations must be satisfied (see, e.g., \cite{Rychkov:2016iqz}):
\begin{equation}
    \label{eq:sct_scaling}
\Phi(b(x) X) = b(x)^{-\Delta_\Phi} \Phi(X)
\end{equation}
where
\begin{equation}
b(x) = \frac{1}{1+2a\cdot x + a^2 x^2}
\end{equation}
is the projective rescaling factor needed to come back to the PS after an $L_s$ Lorentz transformation, i.e. $X' = b(x)\, L_s X$ is a special conformal transformation. This primary condition together with the usual scaling condition constrain the neural network architecture.
Conversely, if one acts on a primary $\Phi$ with derivatives, the associated descendents are no longer homogeneous with respect to special conformal transformations as in \eqref{eq:sct_scaling}. In our non-unitary example of section \ref{sec:NonUnitary}, there are a finite number of descendents, but in general one will have an infinite tower, as in our example of Section \ref{sec:Unitary}.

We emphasize that since we only use the Lorentzian theory as a tool to obtain conformal fields, it is very weakly constrained compared to standard Lorentz invariant field theories. For instance, though it is $SO(D+1,1)$ Lorentz-invariant, it need not be translation invariant, since translation invariance in the CFT on the Poincar\'e section is inherited from Lorentz invariance on the embedding space. It also does not require a well-behaved Hilbert space of states or unitary time evolution, though one could aim to satisfy these additional restrictions with appropriate engineering. Such weak constraints on the Lorentzian theory are to be expected: this is data used to define a Euclidean CFT on $\bR^D$, which in general only requires the conformal invariance inherited from the Lorentz group. In general, these conformal field theories need not be unitary or have a stress-energy tensor, though in some cases they will satisfy the unitarity bound and we may deduce information about the stress tensor from $D=4$ conformal block expansions of exact four-point functions.

\vspace{1cm}
In summary, we obtain a CFT on the Poincar\'e section from a Lorentzian theory associated to a homogeneous neural network architecture. There are three crucial properties to ensure:
\begin{enumerate}
\item \textbf{Homogeneity} arising from the choice of architecture.
\item \textbf{Lorentz-invariance} in $D+2$ dimensions from an appropriately chosen $P(\Theta)$.
\item \textbf{Finiteness.} The correlators must be well-defined.
\end{enumerate}
Obtaining homogeneity and formal Lorentz-invariance is often straightforward, requiring only a  careful choice of architecture and $P(\Theta)$, but ensuring that the correlators are well-defined is a non-trivial task; e.g., unwise choices that we will review can cause them to diverge everywhere. Ensuring finiteness is the main technical challenge in the construction.

\begin{figure}
\centering
\begin{tikzpicture}
    \node[anchor=west] at (2,0) {\begin{minipage}{5cm} \centering
        $SO(D+2)$-symmetric Homogeneous Theory \\  on $\mathbb{R}^{D+2}$
    \end{minipage}};
    
    \draw[->, thick] (8.2,0) -- (10.2,0);
    \node[anchor=west] at (8.6,0.4) {Wick};
    
    \node[anchor=west] at (11,0) {\begin{minipage}{5cm} \centering
        $SO(D+1,1)$-symmetric Homogeneous Theory \\ on $\mathbb{R}^{D+1,1}$
    \end{minipage}};

    \draw[->, thick] (13.5,-1.4) -- (13.5,-3.4);
    \node[anchor=west] at (12.8,-2.2) {\begin{minipage}{3cm} \centering 
    Restrict
    \end{minipage}};

    \node[anchor=west] at (11,-4.4) {\begin{minipage}{5cm} \centering
        CFT on $\bR^D$
    \end{minipage}};
\end{tikzpicture}
\caption{A construction technique pursued in Section \ref{sec:NNCFT}, where Lorentzian correlators are obtained via Wick rotation of analytically computed correlators of a rotationally invariant theory on $\bR^{D+2}$. See Section \ref{sec:other_treatments} for other treatments of the Lorentzian theory.}
\label{fig:RDp2}
\end{figure}
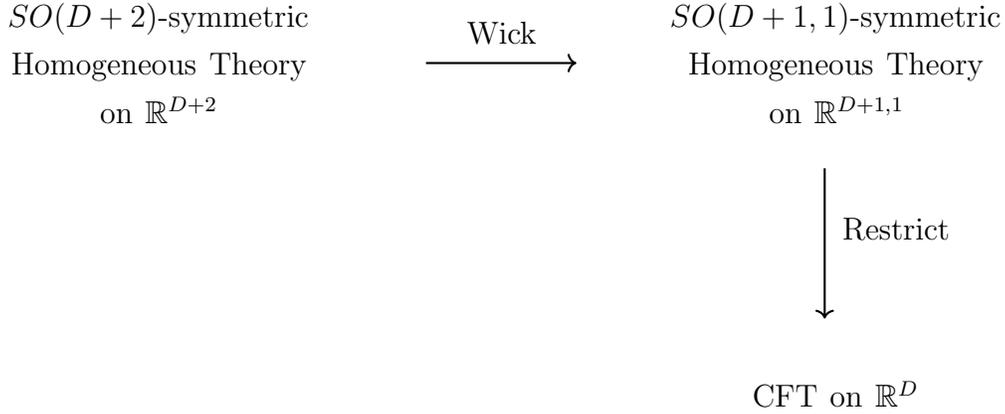

In this section we present simple analytic results that exemplify our CFT construction. One general technical difficulty is obtaining the correlators of the Lorentzian theory, but in this section we sidestep the difficulty by instead computing the correlators of a rotationally invariant Euclidean theory on $\bR^{D+2}$ and performing a Wick rotation. The relationship between the theories with rotation, Lorentz, and conformal symmetry is presented in Figure \ref{fig:RDp2}. In Section \ref{sec:other_treatments} we will present other treatments and subtleties of the Lorentzian correlators. In this section, we will also explain how to get a free theory at large $N$, how to get recursive CFTs from deep neural networks, and will comment on potential numerical approaches.

\subsection{Exactly Solvable Non-Unitary Theories}\label{sec:NonUnitary}

As a first example, consider the Euclidean theory on $\bR^{D+2}$ defined by the Euclidean field
\begin{equation}
    \Phi_E(X) = \left(\Theta \cdot X\right)^{-\Delta}, \qquad P(\Theta) \,\,\, \text{rotationally invariant},
\end{equation}
where we have used the Euclidean dot product.
Since $P(\Theta)$ is rotationally invariant, so is the associated neural network field theory defined by 
\begin{equation}\label{eq:nonunitary_partition_function}
Z_E[J] = \int D\Theta \, P(\Theta)\, e^{\int d^{D+2} X\, J(X) \Phi_E(X)}.
\end{equation}
We have a different field for each $P(\Theta)$ and $\Delta$. Higher fields may be formed by taking a product.

For the sake of illustration we restrict to the case $\Delta = -1$ and our field is 
\begin{equation}\label{eq:nonunitary_field}
    \Phi_E(X) = \Theta \cdot X.
\end{equation}
The Euclidean two-point function is $\bk{\Phi_E(X) \Phi_E(Y)}$ and evaluates to
\begin{equation}
G_E^{(2)}(X_1,X_2) = X_1\cdot X_2
\end{equation}
where we have fixed the normalization by demanding the second moment $\mu_2 := \bk{\Theta_i^2}=D+2$ (no sum) and have Euclidean metric dot product. If we have a vanishing third moment, i.e., $\mu_3=\langle \Theta_1 \Theta_2 \Theta_3\rangle=0$, then the three-point function $G_E^{(3)}(X_1,X_2,X_3)=0$. The four-point function is given by
\begin{equation}
    G_E^{(4)}(X_1,X_2,X_3,X_4) = \sum_{i,j,k,l=1}^D \bk{\Theta_i\Theta_j\Theta_k\Theta_l}\, X_{1i} X_{2j} X_{3k} X_{4l}.
\end{equation}
Rotational symmetry gives the moment tensor $\mu_{ijkl} = \bk{\Theta_i\Theta_j\Theta_k\Theta_l}= \frac{\mu_4}{3}(\delta_{ij}\delta_{kl} + \delta_{ik}\delta_{jl} + \delta_{il}\delta_{jk})$ where $\mu_4 := \bk{\theta_i^4}$ (no sum) is a diagonal element. The four-point function is then
\begin{equation}
    G_E^{(4)}(X_1,X_2,X_3,X_4) = \frac{\mu_4}{3}\left[(X_1\cdot X_2) \, (X_3 \cdot X_4) + \text{perms}\right],
\end{equation}
and the higher correlators may be computed in a similar manner, exactly.

We pass to the Lorentzian theory on $\bR^{D+1,1}$ by Wick rotation, which yields Lorentzian two-point and four-point functions 
\begin{align}
G^{(2)}(X_1,X_2) &= X_1\cdot X_2 \\ G^{(4)}(X_1,X_2,X_3,X_4) &= \frac{\mu_4}{3}\left[(X_1\cdot X_2) \, (X_3 \cdot X_4) + \text{perms}\right],
\end{align}
where here (and whenever there is not a subscript $E$ on $G$) the dot product is to be interpreted in the $(D+2)$-dimensional mostly-plus Minkowski metric to match canonical embedding formalism notation. Restricting to the Poincar\'e section, the CFT correlators are 
\begin{align}\label{eq:delta-1_g2_g4}
G^{(2)}(x_1,x_2)&= x_{12}^2 \\
G^{(4)}(x_1,x_2,x_3,x_4)&= \frac{\mu_4}{3}\left[x_{12}^2 x_{34}^2 + x_{13}^2 x_{24}^2 + x_{14}^2 x_{23}^2 \right],
\end{align}
where, e.g., $x_{12}^2 = (x_1-x_2)^2$. Putting in a canonical cross-ratio form, we have
\begin{align}
G^{(4)}(x_1,x_2,x_3,x_4)&= g(u,v) \, x_{12}^2 x_{34}^2,
\end{align}
where 
\begin{equation}
    g(u,v) =\frac{\mu_4}{3} \left(1+\frac{1}{u}+\frac{v}{u}\right), \qquad u=\frac{x_{12}^2 x_{34}^2}{x_{13}^2x_{24}^2}, \qquad v=\frac{x_{14}^2 x_{23}^2}{x_{13}^2x_{24}^2}.
\end{equation}
We see that the $(u,v)$ dependence of $g(u,v)$ matches that of a free $\Delta = -1$ theory, 
except for the coefficient that renders the theory interacting since the connected four-point function is non-zero. However, in the case that the rotationally invariant density $P(\Theta)$ is a multivariate Gaussian, we have $\mu_4=3\mu_2=3$ and the theory is a generalized free field CFT with $\Delta = -1$. See appendix \ref{app:non-local} for a discussion on non-locality.

Of course, $g(u,v)$ satisfies the crossing constraints \eqref{eq:crossing}, as required by conformal symmetry. Crossing is automatic because our correlators are conformal by construction. This should be contrasted with the bootstrap program where one is searching for CFTs, and therefore crossing symmetry imposes a non-trivial constraint that must be satisfied.

We will now use the four-point function $G^{(4)}_{\Phi}$ to constrain the spectrum of $\phi\times\phi$ OPE in four dimensions. Expanding $g(u,v)$ in the $(z,x)$-variables introduced in \ref{eq:zx-variables},
\begin{equation}
    g(z,\Bar{z})=\frac{\mu_4}{3}\left(\frac{2}{z\Bar{z}}-\frac{1}{z}-\frac{1}{\Bar{z}}+2\right)
\end{equation}
The CBD contains exactly one primary $\mathcal{O}_{0,-2}$ other than the identity and can be written as
\begin{equation}\label{eq:phi_guv}
    g(z,\Bar{z})=\frac{\mu_4}{3}\left(2 g_{-2,0}(z,\Bar{z}) +\frac{4}{3}g_{0,0}(z,\Bar{z})\right)
\end{equation}
where
\begin{equation}
    \begin{split}
        & g_{0,0}(z,\Bar{z})= 1, \\
        & g_{-2,0}(z,\Bar{z})= \frac{1}{z\Bar{z}}-\frac{1}{2z}-\frac{1}{2\Bar{z}}+\frac{1}{3}
    \end{split}
\end{equation}
Since we do not see a contribution from an operator $\mathcal{O}_{4,2}$ to $g(u,v)$, we infer that this theory does not have a local stress tensor. We have also seen that a primary of dimension $-2$ exists in the theory, which should be identified with $\Phi^2$.

We therefore wish to study $\Phi^2$. The two-point and three-point functions are given by
\begin{equation}
\begin{aligned}
    & G^{(2)}_{\Phi^2\Phi^2}(X_1,X_2)=\frac{2\mu_4}{3} (X_1\cdot X_2)^2 := \frac{2\mu_4}{3} (x_{12})^4 \\
    & G^{(3)}_{\Phi^2\Phi^2\Phi^2}(X_1,X_2,X_3)= \frac{4\mu_6}{15} (X_1\cdot X_2) (X_1\cdot X_3) (X_2\cdot X_3) = \frac{4\mu_6}{15} (x_{12})^2 (x_{13})^2 (x_{23})^2
\end{aligned}
\end{equation}
consistent with \ref{eq:2pt-scalar} and \ref{eq:3pt-scalar}, whereas the three-point function between two $\Phi$s and one $\Phi^2$ is given by
\begin{equation}
    G^{(3)}_{\Phi\Phi\Phi^2}(X_1,X_2,X_3)=2\mu_4(X_1\cdot X_3) (X_2\cdot X_3) := \frac{2\mu_4}{3} (x_{13})^2 (x_{23})^2.
\end{equation}
Using the same technique as $G^{(4)}_\Phi$, we may compute the four-point function $G^{(4)}_{\Phi^2}$ in the CFT by beginning on $\bR^{D+2}$, Wick rotating, and restricting to the PS. The eighth moment tensor
\begin{equation}
    \bk{\theta_{a}\dots \theta_h} = \frac{\mu_8}{105} \left(\delta_{ab} \delta_{cd} \delta_{ef} \delta_{gh} + 104\, \text{perms}\right)
\end{equation}
appears in the calculation where $\mu_8 = \bk{\theta_i^8}$ (no sum) is the diagonal element. Upon Wick rotation and restriction, only two of the possible five diagram topologies survive, giving in the CFT on $\bR^D$
\begin{equation}
G^{(4)}_{\Phi^2} = \frac{\mu_8}{105}\left[4(x_{12}^4x_{34}^4+x_{13}^4x_{24}^4+x_{14}^4x_{23}^4) + 16(x_{12}^2 x_{23}^2x_{34}^2 x_{14}^2+x_{12}^2 x_{24}^2x_{34}^2 x_{13}^2+x_{13}^2 x_{23}^2x_{24}^2 x_{14}^2) \right],
\end{equation}
which can be written in the $s$-channel as 
$G^{(4)}_{\Phi^2} = x_{12}^4x_{34}^4\, g_{\Phi^2}(u,v)$ with
\begin{equation}
g_{\Phi^2}(u,v) = \frac{\mu_8}{105}\left[4\left(1+\frac{1}{u^2}+\frac{v^2}{u^2}\right) + 16\left(\frac{v}{u} + \frac{1}{u} + \frac{v}{u^2}\right) \right].
\end{equation}\label{eq:phi2guv}
We can write \ref{eq:phi2guv} as a sum of the four-dimensional conformal blocks as
\begin{equation}\label{eq:phi2_guv}
    g_{\Phi^2}(z,\Bar{z})=\frac{4\mu_8}{105}\left( 6g_{-4,0}(z,\Bar{z})+ \frac{48}{5} g_{-2,0}(z,\Bar{z}) +\frac{5}{2}g_{0,0}(z,\Bar{z})\right)
\end{equation}
where 
\begin{equation}
    g_{-4,0}(z,\Bar{z})=\frac{1}{z^2\Bar{z}^2}-\frac{1}{\Bar{z}^2z}+\frac{1}{6}\frac{1}{\Bar{z}^2}-\frac{1}{z^2\Bar{z}}+\frac{16}{15}\frac{1}{z\Bar{z}}-\frac{1}{5}\frac{1}{\Bar{z}}+\frac{1}{6}\frac{1}{z^2}-\frac{1}{5}\frac{1}{z}+\frac{1}{20}.
\end{equation}
is a conformal block associated with a scalar primary of dimension $-4$. Similarly, we are able to establish the existence of a primary $\Phi^8$ via $g_{\Phi^4}$, and expect that the pattern persists up to high powers.

\subsection{Conformal Block / OPE Matching in Neural Network Approach}

The coefficients of $g_{\mathcal{O}}$'s in~(\ref{eq:phi_guv}) and in~(\ref{eq:phi2_guv}) should match the results from $\Phi\times\Phi$ OPE and $\Phi^2\times\Phi^2$ OPE respectively.  
More precisely, the constant in front of $g_{\mathcal{O}}$ in the CBD should be equal to $\lambda_{\Phi^k\Phi^k\mathcal{O}}^2$ which in turn is the constant coefficient of $\frac{\langle \Phi^k(X|_\text{PS})\Phi^k(Y|_\text{PS}) \mathcal{O}(Z|_\text{PS}) \rangle^2}{\langle \mathcal{O}(X|_\text{PS})\mathcal{O}(Y|_\text{PS}) \rangle}$. These 3-point and 2-point functions can be computed via embedding formalism using~(\ref{eq:Gx_from_GX}) for any field $\mathcal{O}$ that shows up in the $\Phi^k\times \Phi^k$ OPE. One can immediately see that $\lambda_{\Phi^k\Phi^k\mathcal{O}}^2$ is precisely the constant coefficient of $\frac{\langle \Phi^k(X)\Phi^k(Y) \mathcal{O}(Z) \rangle^2}{\langle \mathcal{O}(X)\mathcal{O}(Y) \rangle}$ when restricted to the null cone of the embedding space, hence the CBD coefficients can be computed from embedding space correlators restricted to the null cone.

Before we begin, we'd like to discuss some essential aspects of the OPE in our neural network approach to conformal fields. Since the OPE In the CFT, we have  The self-OPE of an operator in the embedding space is of the form 

\begin{equation}
    \cO_\Theta(X)\times \cO_{\Theta}(Y) \sim \sum_{\cO'} \lambda_{\cO\cO\cO'}\cO'_\Theta(Y)
\end{equation}
where we use $\sim$ to denote that this is an operator relation for the conformal family (where we have suppressed the descendants), and the subscript $\Theta$ emphasizes that in our formalism, an operator $\mathcal{O}_{\Theta}(X)$ can in general depend on parameters $\Theta$. In the example of this section, where we study operators $\Phi_\Theta(X) = \Theta\cdot X$ and $\Phi_\Theta^2$, we see in both cases that the operators are homogeneous in $\Theta$. This is true in much more general neural networks\footnote{For instance, those where $\Theta$ are the subset of parameters responsible for absorbing rotations or Lorentz transformations. If it does so linearly, then homogeneity of $X$ ensures homogeneity of $\Theta$, and in particular any neural net of the form $f_\varphi(\Theta \cdot X)$ homogeneous in $X$, potentially with more parameters $\varphi$, is of this type.}. A consistent OPE requires
\begin{equation}
\label{eqn:OPE_theta_scaling}
\Delta^{\Theta}_{\cO'} = 2\Delta^{\Theta}_\cO, \qquad \forall \cO'\in \Phi \times \Phi,
\end{equation}
where $\Delta^\Theta_\cO$ is the scaling dimension of $\cO$ in $\Theta$\footnote{This should not be confused with the scaling dimension $\Delta_{\mathcal{O}}$ of operator $\mathcal{O}$.}, in order to ensure homogeneity of $\Theta$ on both sides of the equation, and that $\lambda_{\cO\cO\cO'}$ does not depend on $\Theta$.
The independence of the OPE coefficients on $\Theta$ follows from the fact that the architecture of the operator is a map from parameters to fields, and the OPE should be the same for all values of the parameters. Put differently, this is how the OPE is an operator equation in our context, valid in any appropriate expectation. The requirement \eqref{eqn:OPE_theta_scaling} constrains $\cO'$, and trivially generalizes for $\cO_1\times \cO_2$ OPE in which case we require:
\begin{equation}\label{eq:Theta_homogeneity}
    \Delta_{\cO'}^\Theta = \Delta_{\cO_1}^\Theta + \Delta_{\cO_2}^\Theta, \qquad \forall \cO'\in \cO_1 \times \cO_2
\end{equation}
which we will call $\Theta$-homogeneity condition.

To read off the coefficients of CBD from the corresponding OPE, we first define:
\begin{equation}\label{eq:ansatz-field}
    \mathcal{O}_{m,n}(X):= (\Theta\cdot\Theta)^m (\Theta\cdot X)^n,
\end{equation} 
in the embedding space with $\Delta^{\Theta}_{\mathcal{O}_{m,n}} = 2m+n$. Since $\Delta_{\mathcal{O}} = -n$, by the embedding formalism it restricts to a field in the physical spacetime $\mathbb{R}^D$ with conformal dimension $\Delta = -n$. Therefore, there is an apparent ambiguity in determining which $\mathcal{O}_{m,n}(X)$ restricts to the conformal field with dimension $-n$ that appears in the CBD. However, we will see that applying~(\ref{eqn:OPE_theta_scaling}) to the CBD of $g_{\Phi^k}$ uniquely determines which $\mathcal{O}_{m,n}(X)$ restricts to the conformal field with dimension $-n$.

\vspace{.5cm}
\noindent \textbf{CBD coefficients of $G_{\Phi\Phi\Phi\Phi}$}

Let us first attempt to compute the coefficients of~(\ref{eq:phi_guv}) from $\Phi\times\Phi$ OPE with the help of~(\ref{eqn:OPE_theta_scaling}). From now on, for simplicity, by $G_{\cO_1\cO_2\cdots}$ we mean the VEV $\langle \cO_1(X_1) \cO_1(X_2) \cdots \rangle$ in the embedding space and similarly for the correlators in the physical spacetime. We first look at the coefficient of $g_{-2,0}$, for which we compute the following correlators:
\begin{equation}
	\begin{split}
		G_{\Phi\Phi\Phi^2}(X,Y,Z) &= \mathbb{E}[\Theta_i\Theta_j\Theta_{k_1}\Theta_{k_2}] X^iY^jZ^{k_1}Z^{k_2} =  \frac{2\mu_4}{3} (X\cdot Z)(Y\cdot Z), \\ 
	   G_{\Phi^2\Phi^2}(X,Y) &= \mathbb{E}[\Theta_{i_1}\Theta_{j_1}\Theta_{i_2}\Theta_{j_2}] X^{i_1}X^{j_1}Y^{i_2}Y^{j_2} = \frac{2\mu_4}{3} (X\cdot Y)^2
	\end{split}
\end{equation}
presuming that the $\Delta = -2$ field in the $\Phi\times\Phi$ OPE is given by $\Phi^2 \equiv \mathcal{O}_{0,2}$, which is the only $\cO_{m,2}$ allowed by the homogeneity condition~\eqref{eqn:OPE_theta_scaling}. Thus the coefficient of $g_{-2,0}$ can be computed to be $\left(\frac{2}{3}\mu_4\right)^2/\left(\frac{2}{3}\mu_4\right)  = \frac{2}{3}\mu_4$ which matches that in~(\ref{eq:phi_guv}). For the coefficient of $g_{0,0}$, we need correlators involving an $\cO_{m,0}$, which is fixed to $\cO_{1,0}$ by the condition \eqref{eqn:OPE_theta_scaling}. Thus we compute the following correlators:
\begin{equation}
	\begin{split}
		G_{\Phi\Phi \mathcal{O}_{1,0}}(X,Y,Z) &= \delta^{i_1j_1} \mathbb{E}[\Theta_{i_1}\Theta_{j_1}\Theta_k\Theta_j]X^k Y^j = \frac{\mu_4}{3}((D+2)+2) X\cdot Y,\\
		G_{\mathcal{O}_{1,0} \mathcal{O}_{1,0}}(X,Y) &= \delta^{i_1j_1} \delta^{i_2j_2} \mathbb{E}[\Theta_{i_1}\Theta_{j_1} \Theta_{i_2}\Theta_{j_2}] =  \frac{\mu_4}{3}((D+2)^2 + 2(D+2))
	\end{split}
\end{equation}
The coefficient of $g_{0,0}$ when $D = 4$ can be computed to be $\frac{\mu_4}{3} \frac{((D+2)+2)^2}{(D+2)^2 + 2(D+2)} = \frac{4}{3}\frac{\mu_4}{3}$ which again matches that in~(\ref{eq:phi_guv}). Therefore,~(\ref{eq:phi_guv}) is reproduced from OPE.

\vspace{.5cm}
\noindent \textbf{CBD coefficients of $G_{\Phi^2\Phi^2\Phi^2\Phi^2}$}

We now attempt to compute the coefficients of~(\ref{eq:phi2_guv}) from $\Phi^2\times\Phi^2$ OPE. We first look at the coefficient of $g_{-4,0}$. By \eqref{eqn:OPE_theta_scaling}, the only $\cO_{m,4}$ that can be the $\Delta = -4$ operator in the $\Phi^2\times\Phi^2$ OPE is fixed by~(\ref{eq:Theta_homogeneity}) to $\Phi^4\equiv \mathcal{O}_{0,4}$. Hence we compute the following correlators:
\begin{equation}
	\begin{split}
		G_{\Phi^2\Phi^2 \Phi^4}(X,Y,Z) &= \mathbb{E}[\Theta_{i_1}\Theta_{i_2}\Theta_{j_1}\Theta_{j_2} \Theta_{k_1}\Theta_{k_2}\Theta_{k_3}\Theta_{k_4}] X^{i_1}X^{i_2} Y^{j_1}Y^{j_2} Z^{k_1}Z^{k_2}Z^{k_3}Z^{k_4} \\
        &= \frac{24\mu_8}{105} (X\cdot Z)^2 (Y\cdot Z)^2, \\
		G_{\Phi^4 \Phi^4}(X,Y) &= \mathbb{E}[\Theta_{i_1}\Theta_{i_2}\Theta_{i_3}\Theta_{i_4} \Theta_{j_1}\Theta_{j_2}\Theta_{j_3}\Theta_{j_4}] X^{i_1}X^{i_2}X^{i_3}X^{i_4} Y^{j_1}Y^{j_2}Y^{j_3}Y^{j_4} \\
        &= \frac{24\mu_8}{105} (X\cdot Y)^4\,.
	\end{split}
\end{equation}
The coefficients of $g_{-4,0}$ can be computed to be $\left(\frac{\mu_8}{105} \times 24\right)^2/\left(\frac{\mu_8}{105} \times 24\right) = \frac{24\mu_8}{105}$ which matches that in~(\ref{eq:phi2_guv}). For the coefficient of $g_{-2,0}$, the only $\cO_{m,n}$ that can be the $\Delta = -2$ operator in the $\Phi^2\times\Phi^2$ OPE is fixed by~(\ref{eq:Theta_homogeneity}) to $\mathcal{O}_{1,2}$, Hence we compute the following correlators:
\begin{equation}
	\begin{split}
		G_{\Phi^2\Phi^2 \mathcal{O}_{1,2}}(X,Y,Z) &= \delta^{ij}  \mathbb{E}[\Theta_{i}\Theta_{j} \Theta_{j_1}\Theta_{j_2} \Theta_{k_1}\Theta_{k_2} \Theta_{l_1}\Theta_{l_2}] X^{j_1}X^{j_2} Y^{k_1}Y^{k_2} Z^{l_1}Z^{l_2} \\
        &= \frac{\mu_8}{105} (8(D+2) + 48) (X\cdot Y) (X\cdot Z) (Y\cdot Z), \\
		G_{\mathcal{O}_{1,2} \mathcal{O}_{1,2}}(X,Y) &= \delta^{i_1j_1} \delta^{i_2j_2} \mathbb{E}[\Theta_{i_1}\Theta_{j_1} \Theta_{i_2}\Theta_{j_2} \Theta_{k_1}\Theta_{k_2} \Theta_{l_1}\Theta_{l_2}] X^{k_1}X^{k_2}Y^{l_1}Y^{l_2} \\
        &= \frac{\mu_8}{105} (2(D+2)^2 + 20(D+2) + 48) (X\cdot Y)^2\,.
	\end{split}
\end{equation}
When $D = 4$, the coefficient of $g_{-2,0}$ can be computed to be
\begin{align*}
    \frac{\mu_8}{105} \frac{(8(D+2) + 48)^2}{2(D+2)^2 + 20(D+2) + 48} = \frac{192}{5} \frac{\mu_8}{105}
\end{align*}
which matches that in~(\ref{eq:phi2_guv}). Finally, for the coefficient of $g_{0,0}$,  the only $\cO_{m,n}$ that can be the $\Delta = 0$ operator in the $\Phi^2\times \Phi^2$ OPE is fixed by~(\ref{eq:Theta_homogeneity}) to $\mathcal{O}_{2,0}$, and we compute the following correlators:
\begin{equation}
	\begin{split}
        G_{\Phi^2\Phi^2\mathcal{O}_{2,0}}(X,Y,Z) &= \frac{\mu_8}{105} \delta^{i_1j_1} \delta^{i_2j_2} \mathbb{E}[\Theta_{i_1}\Theta_{j_1}\Theta_{i_2}\Theta_{j_2} \Theta_{k_1} \Theta_{k_2} \Theta_{l_1} \Theta_{l_2}] X^{k_1} X^{k_2} Y^{l_1} Y^{l_2} \\
        &= \frac{\mu_8}{105} (2(D+2)^2 + 20(D+2) + 48) (X\cdot Y)^2, \\
		G_{\mathcal{O}_{2,0}\mathcal{O}_{2,0}}(X,Y) &= \frac{\mu_8}{105}\delta^{i_1j_1}\delta^{i_2j_2}\delta^{k_1l_1}\delta^{k_2l_2}\mathbb{E}[\Theta_{i_1}\Theta_{j_1}\Theta_{i_2}\Theta_{j_2}\Theta_{k_1}\Theta_{l_1}\Theta_{k_2}\Theta_{l_2}] \\
        &= \frac{\mu_8}{105}((D+2)^4 + 12 (D+2)^3 + 44 (D+2)^2 + 48 (D+2))\,.
	\end{split}
\end{equation}
Therefore, when $D = 4$ the coefficient of $g_{0,0}$ can be computed to be
\begin{align*}
    \frac{\mu_8}{105} \frac{(2(D+2)^2 + 20(D+2) + 48)^2}{(D+2)^4 + 12(D+2)^3 + 44(D+2)^2 + 48(D+2)} = \frac{10\mu_8}{105}
\end{align*}
which matches with that in~(\ref{eq:phi2_guv}).

We summarize the above results in Table~\ref{tab:dimDelta_OPE}.
\begin{table}[h]
    \centering
    \begin{tabular}{c|c|c}
        & $\Phi\times\Phi$ & $\Phi^2\times\Phi^2$ \\
        \hline
        \hline
        $\Delta = 0$ & $\mathcal{O}_{1,0}$ & $\mathcal{O}_{2,0}$ \\
        \hline
        $\Delta = -2$ & $\mathcal{O}_{0,2}$ & $\mathcal{O}_{1,2}$ \\
        \hline
        $\Delta = -4$ & - & $\mathcal{O}_{0,4}$ \\
    \end{tabular}
    \caption{The ansatz for dimension $\Delta$ fields in $\Phi\times\Phi$ OPE and $\Phi^2\times\Phi^2$ OPE that lead to consistent coefficients of the CBD~(\ref{eq:phi_guv}) and~(\ref{eq:phi2_guv}), respectively.}
    \label{tab:dimDelta_OPE}
\end{table}
We note that in order to correctly match the coefficients of corresponding CBD, the dimension $\Delta$ fields in $\Phi^k\times\Phi^k$ OPE have to come from the ansatz given in \ref{eq:ansatz-field}. It is readily seen that this is a manifestation of~(\ref{eqn:OPE_theta_scaling}), e.g., in $\Phi\times\Phi$ OPE the $\Delta = -2$ operator is $\mathcal{O}_{0,2}$ since $\Delta^\Theta_{\mathcal{O}_{0,2}} = 2 = 2\times \Delta^\Theta_{\Phi}$ whereas in $\Phi^2\times\Phi^2$ OPE the $\Delta = -2$ operator is $\mathcal{O}_{1,2}$ since $\Delta^\Theta_{\mathcal{O}_{1,2}} = 4 = 2\times \Delta^\Theta_{\Phi^2}$. This, however, forces us to pick different $\mathcal{O}_{m,n}$'s to be the $\Delta=0$ operators in the $\Phi\times\Phi$ and $\Phi^2\times\Phi^2$ OPEs respectively.

\vspace{0.5cm}
\noindent \textbf{CBD coefficients of $G_{\Phi\Phi\Phi^2\Phi^2}$}

We further check our claim against other types of correlators. Let us first consider the correlator $G_{\Phi\Phi\Phi^2\Phi^2}(X_1,X_2,X_3,X_4)$ where $\Phi(X)=\Theta \cdot X$ and $\Phi^2(X)=(\Theta\cdot X)^2$. We have:
\begin{equation}\label{eq:mixed-phi-phi2}
\begin{aligned}
    G_{\Phi\Phi\Phi^2\Phi^2}(X_1,X_2,X_3,X_4) & =\frac{2\mu_6}{15} (X_1\cdot X_2)(X_3 \cdot X_4)^2 \left[1+2 \left(\frac{(X_1\cdot X_3) (X_2 \cdot X_4)}{(X_1\cdot X_2) (X_3 \cdot X_4)} \right.\right.\\
    &\qquad \left.\left.+ \frac{(X_1\cdot X_4) (X_2 \cdot X_3)}{(X_1\cdot X_2) (X_3 \cdot X_4)} \right)\right]\\
    & = \frac{2\mu_6}{15} (x_{12})^2 (x_{34})^4 \left[1+2\left(\frac{1}{u}+ \frac{v}{u}\right)\right]\\
    &= (x_{12})^2 (x_{34})^4 g(u,v),
\end{aligned}
\end{equation}
given which we define:
\begin{equation}
    \label{eq:g_mixed}
    g(u,v) := \frac{2\mu_6}{15}\left[1+2\left(\frac{1}{u}+ \frac{v}{u}\right)\right]=\frac{2\mu_6}{15}\left( 3 - \frac{2}{\Bar{z}} - \frac{2}{z} + \frac{4}{z\Bar{z}} \right).
\end{equation}
It is not hard to compute the CBD of $g(u,v)$ to be:
\begin{equation}
    \label{eq:gmixed_cbd}
    g(z,\Bar{z})= \frac{2\mu_6}{15} \left(4 g_{-2,0}(z,\Bar{z}) + \frac{5}{3} g_{0,0}(z,\Bar{z}) \right),
\end{equation}
which is consistent with the expectation that $G_{\Phi\Phi\Phi^2\Phi^2}$ can contain only those operators that are present in the spectra of both $\Phi\times\Phi$ OPE and $\Phi^2\times\Phi^2$ OPE, which in this case are the dimension-$0$ and dimension-$(-2)$ fields.

We now compute the CBD coefficients via OPE using~(\ref{eq:Theta_homogeneity})
Here we are concerned with both $\Phi \times \Phi$ OPE and $\Phi^2 \times \Phi^2$ OPE, therefore the two operators of equal scaling dimension coming from the two OPEs can differ in their $\Theta$-degrees. This makes it more subtle as we will discuss below. Recall we have the correlators:
\begin{equation}
\begin{split}
    & G_{\Phi\Phi\mathcal{O}_{1,0}}(X,Y,Z)=\frac{\mu_4}{3}(D+4) (X\cdot Y) = \frac{8\mu_4}{3} (X\cdot Y)\\
    & G_{\Phi^2\Phi^2\mathcal{O}_{2,0}}(X,Y,Z)=\frac{\mu_8}{105}(2(D+2)^2 + 20(D+2) + 48) (X\cdot Y)^2=\frac{240 \mu_8}{105} (X\cdot Y)^2\\
    & G_{\mathcal{O}_{1,0} \mathcal{O}_{1,0}}(X,Y)=  \frac{\mu_4}{3}((D+2)^2 + 2(D+2))=\frac{48 \mu_4}{3}=c_{\mathcal{O}_{1,0} \mathcal{O}_{1,0}}\\
    & G_{\mathcal{O}_{2,0}\mathcal{O}_{2,0}}(X,Y) = \frac{\mu_8}{105}((D+2)^4 + 12 (D+2)^3 + 44 (D+2)^2 + 48 (D+2)) = \frac{5760 \mu_8}{105}=c_{\mathcal{O}_{2,0}\mathcal{O}_{2,0}}
\end{split}
\end{equation}

Unfortunately, we were not able to show that this mixed correlator matches the conformal block decomposition. In performing the $\Phi\times \Phi$ and $\Phi^2 \times \Phi^2$ OPE in the correlator, two-point functions of operators of common dimension but mixed $\Theta$-degree appear, such as
\begin{equation}
    G_{\mathcal{O}_{2,0} \mathcal{O}_{1,0}}(X,Y)= \frac{\mu_6}{15} ((D+2)^3+6(D+2)^2+8(D+2))=\frac{480 \mu_6}{15}=c_{\mathcal{O}_{2,0} \mathcal{O}_{1,0}},
\end{equation}
as does the inverse two-point function in a fixed dimension sector, $G^{IJ}$. This inverse two-point function generally mixes $\Theta$-degree and therefore has multiple moments, $\mu_4, \mu_6, \mu_8$, etc., that in general make the required matches with \eqref{eq:gmixed_cbd} (which only depends on $\mu_6$) impossible unless the moments are further constrained. However, this is not strictly required by conformal invariance in our model, and therefore we see that the OPE / CBD matching does not happen in general for this mixed correlator, despite being conformally invariant.

\vspace{0.5cm}
\noindent \textbf{CBD coefficients of $G_{\Phi\Phi^2\Phi\Phi^2}$}

We now compute the CBD coefficients of 4-point function $G_{\Phi\Phi^2\Phi\Phi^2}(X_1,X_2,X_3,X_4)$. We have:
\begin{equation}\label{eq:G1212}
	\begin{split}
		G_{\Phi\Phi^2\Phi\Phi^2}(X_1,X_2,X_3,X_4) &= \mathbb{E}[\Theta_i\Theta_{j_1}\Theta_{j_2}\Theta_k\Theta_{l_1}\Theta_{l_2}]X_1^iX_2^{j_1}X_2^{j_2}X_3^kX_4^{l_1}X_4^{l_2} \\
		&= \frac{2\mu_6}{15}\left( 2(X_1\cdot X_2)(X_3\cdot X_4)(X_2\cdot X_4) + 2(X_1\cdot X_4)(X_2\cdot X_3)(X_2\cdot X_4) \right.\\
        &\qquad\quad \left.+ (X_1\cdot X_3) (X_2\cdot X_4)^2 \right)\,.
	\end{split}
\end{equation}
Further restricting to PS,~(\ref{eq:G1212}) becomes:
\begin{equation}
    \begin{split}
        G_{\phi\phi^2\phi\phi^2} &= \frac{2\mu_6}{15} x_{12}^3 x_{34}^3 \frac{x_{24}}{x_{13}}  \left( 2\frac{x_{13}x_{24}}{x_{12}x_{34}} + 2\frac{x_{14}^2x_{23}^2x_{13}x_{24}}{x_{12}^3x_{34}^3} + \frac{x_{13}^3x_{24}^3}{x_{12}^3x_{34}^3} \right) \\
	&= \frac{2\mu_6}{15} x_{12}^3 x_{34}^3 \frac{x_{24}}{x_{13}} \left( 2u^{-1/2} + 2 v u^{-3/2} + u^{-3/2} \right)
    \end{split}
\end{equation}
given which we define:
\begin{equation}\label{eq:FF2FF2_guv}
    g(u,v) = \frac{2\mu_6}{15} \left( 2u^{-1/2} + 2 v u^{-3/2} + u^{-3/2} \right)\,.
\end{equation}
To find the CBD of the above $g(u,v)$, we use the basic function~(\ref{eq:basic_function}) for $\Delta_{12} = \Delta_{34} = 1$ with which it is easy to show that:
\begin{equation}
    g_{-1,0}(z,\Bar{z}) = \frac{1}{(z\Bar{z})^{1/2}},\ g_{-3,0}(z,\Bar{z}) = \frac{3z\Bar{z} - 4\Bar{z} - 4z + 6}{6(z\Bar{z})^{3/2}}\,.
\end{equation}
Therefore, written in $z,\Bar{z}$ variables, the CBD of~(\ref{eq:FF2FF2_guv}) is:
\begin{equation}\label{eq:CBD_FF2FF2}
    g(z,\Bar{z}) = \frac{\mu_6}{15} \left( 6g_{-3,0}(z,\Bar{z}) + 5g_{-1,0}(z,\Bar{z}) \right)\,.
\end{equation}
Note that the dimension $(-1)$ operator and the dimension $(-3)$ operator appear in the CBD as expected.

We are now ready to check our OPE approach against the CBD coefficients of~(\ref{eq:CBD_FF2FF2}). We first attempt to match the coefficient of $g_{-3,0}$. Since $\Delta^\Theta_\Phi+\Delta^\Theta_{\Phi^2} = 3$, by~(\ref{eq:Theta_homogeneity}) the dimension-$(-3)$ operator in $\Phi\times \Phi^2$ OPE has to be $\cO_{0,3}$ as $\Delta^\Theta_{\cO_{0,3}} = 3$. Hence we need to compute the following correlators:
\begin{equation}
	\begin{split}
		G_{\Phi\Phi^2\cO_{0,3}} &= \mathbb{E}[\Theta_i\Theta_{j_1}\Theta_{j_2}\Theta_{k_1}\Theta_{k_2}\Theta_{k_3}] X^i Y^{j_1}Y^{j_2} Z^{k_1}Z^{k_2}Z^{k_3} = \frac{6\mu_6}{15} (X\cdot Z)(Y\cdot Z)^2, \\
		G_{\cO_{0,3}\cO_{0,3}} &= \mathbb{E}[\Theta_{i_1}\Theta_{i_2}\Theta_{i_3}\Theta_{j_1}\Theta_{j_2}\Theta_{j_3}] X^{i_1}X^{i_2}X^{i_3}Y^{i_1}Y^{i_2}Y^{i_3} = \frac{6\mu_6}{15} (X\cdot Y)^3\,.
	\end{split}
\end{equation}
The coefficients of $g_{-3,0}$ can be computed to be $\left(\frac{6\mu_6}{15}\right)^2/\left(\frac{6\mu_6}{15}\right) = \frac{6\mu_6}{15}$ which matches that in~(\ref{eq:CBD_FF2FF2}).

We then look at the coefficient of $g_{-1,0}$. Since $\Delta^\Theta_\Phi+\Delta^\Theta_{\Phi^2} = 3$, by~(\ref{eq:Theta_homogeneity}) the dimension-$(-1)$ operator that appears in the $\Phi\times \Phi^2$ OPE has to be $\cO_{1,1}$ as $\Delta^\Theta_{\cO_{1,1}} = 3$. Hence we need to compute the following correlators:
\begin{equation}
	\begin{split}
		G_{\Phi\Phi^2\cO_{1,1}} &= \delta^{l_1l_2}\mathbb{E}[\Theta_i\Theta_{j_1}\Theta_{j_2}\Theta_k\Theta_{l_1}\Theta_{l_2}] X^i Y^{j_1}Y^{j_2} Z^k = \frac{\mu_6}{15}(2(D+2)+8)(X\cdot Y)(Y\cdot Z)\\
		G_{\cO_{1,1}\cO_{1,1}} &= \delta^{k_1k_2}\delta^{l_1l_2}\mathbb{E}[\Theta_i\Theta_j\Theta_{k_1}\Theta_{k_2}\Theta_{l_1}\Theta_{l_2}] X^i Y^j  = \frac{\mu_6}{15}((D+2)^2+6(D+2)+8)(X\cdot Y)
	\end{split}
\end{equation}
When $D = 4$, this gives the coefficient of $g_{-1,0}$ to be
\begin{equation}
    \frac{\mu_6}{15} \frac{(2(D+2)+8)^2}{(D+2)^2 + 6(D+2) + 8} = \frac{5\mu_6}{15}
\end{equation}
which matches that in~(\ref{eq:CBD_FF2FF2}). This concludes our discussion on the explicit calculation of CBD coefficients via OPE and the application of $\Theta$-homogeneity condition~(\ref{eq:Theta_homogeneity}).

\vspace{.5cm}
\noindent \textbf{$\Phi\times \Phi$ OPE }

Finally, for the sake of concreteness, we wish to write down the $\Phi \times \Phi$ OPE of this theory. Previously we utilized associated data by showing that the coefficients of the CBD of various four-point functions are the square of appropriate three-point coefficients squared divided by two-point normalizations. We wish to see this arise by direct calculation.

The conformal block decomposition of the $\Phi$ four-point function tells us that the OPE is of the form
\begin{equation}
    \Phi(X) \Phi(Y) \sim A(X,Y) \, \Theta \cdot \Theta + B(X,Y) (\Theta \cdot Y)^2
\end{equation}
where we have imposed our $\Theta$-homogeneity condition and remembered that the coefficient may also depend on $X$ and $Y$. Using the OPE in the four-point function and recalled our definitions $\cO_{1,0}(Y) = \Theta \cdot \Theta$ and $\Phi^2(Y) = \Theta \cdot Y$, we have 
\begin{align}
    G^{(4)}(X,Y,Z,W) &= \langle \Phi(X) \Phi(Y) \Phi(Z) \Phi(W)  \rangle \\
    &= A(X,Y) G^{(3)}_{\cO_{10}\Phi(Z) \Phi(W)} + B(X,Y)G^{(3)}_{\Phi^2(Y)\Phi(Z) \Phi(W)}\\
    &= A(X,Y)A(Z,W) G^{(2)}_{\cO_{10}\cO_{10}} + B(X,Y)B(Z,W) G^{(2)}_{\Phi^2(Y)\Phi^2(W)}
\end{align}
where we remember that all expectations are computed as integrals against $P(\Theta)$, and the second (third) equality arises after the application of the OPE once (twice). Matching terms, we find 
\begin{align}
    A(Z,W) &= \frac{G^{(3)}_{\cO_{10}\Phi(Z) \Phi(W)}}{G^{(2)}_{\cO_{10}\cO_{10}}} = \frac{\mu_4}{3 D} (Z\cdot W)^2 \\ 
    B(Z,W) &= \frac{G^{(3)}_{\Phi^2(Y) \Phi(Z) \Phi(W)}}{G^{(2)}_{\Phi^2(Y)\Phi^2(W)}} = \frac{2\mu_4}{3}.
\end{align}
This data is derived for operators with unnormalized two-point function. If one were instead to use normalized operators so $A \cO\cO$ is replaced by $\hat A \hat \cO \hat \cO$, then equality of these terms requires $\hat A = \sqrt{C_{{\cO}{\cO}}} A$, where $C_{{\cO}{\cO}}$ is the normalization of $G^{(2)}_{\cO\cO}$. In terms of these operators, the final OPE becomes 
\begin{equation}
    \Phi(X) \Phi(Y) \sim \,\frac{(\mu_4/3)(D+2)}{\sqrt{(\mu_4/3)D(D+2)}}\, (X\cdot Y)^2 \widehat{\cO_{10}(Y)} + \frac{2\mu_4/3}{\sqrt{2\mu_4/3}} \widehat{\Phi(Y)^2},
\end{equation}
where we have written the OPE coefficients in a way that explicitly matches our previous calculations: their square is the coefficient of appropriate three-point function squared divided by two-point normalization.

\subsection{Free Theory Limit at Large-$N$}
\label{sec:freetheory}

The harbinger of a free theory is that the connected correlation functions satisfy $G^{(2n)}_c=0$ for all $n>2$. In the machine learning literature, a result known as the neural network / Gaussian process (NNGP) correspondence yields this result at large-$N$ due to Central Limit Theorem, where $N$ is associated to some aspect of the architecture such as the width of a deep neural network. In particular, under mild assumptions 
\begin{equation}
    G^{(2n)}_c \propto \frac{1}{N^{n-1}}
\end{equation}
which is sufficient to ensure Gaussianity as $N\to\infty$. In such a limit, we say that neural networks are drawn from a Gaussian process, or generalized free field theory.

We adapt this to the case of our CFT construction. Let us simply consider the following architecture:
\begin{equation}\label{eq:largeN_architecture}
    \varphi(X) = \frac{1}{\sqrt{N}} \sum_{i=1}^N w_i \Phi_i(X).
\end{equation}
where $\{w_i\}\sim P(w)$ i.i.d. and $\Phi_i(X)$ is any collection of neural networks drawn from the same distribution, that is also homogeneous and defines a Lorentz invariant theory via any of the mechanism in this paper. This ensures that $\phi(x)=\Phi|_\text{PS}$ defines a conformal field, which we assume has a canonically normalized two-point function. We assume the following properties of the distribution of each $w_i$:
\begin{align}\label{eq:free-theory-limit-params}
    \bk{w_i^{2k+1}} = 0,\ \bk{w_i^2} = 1,\ \bk{w_i^4} = \gamma^4,\qquad \forall i = 1,\cdots,N,\qquad \forall k\in \mathbb{N}.
\end{align}
The two-point function on $\bR^{D+1,1}$ is
\begin{equation}\label{eq:G2_EE}
    G^{(2)}(X_1,X_2) = \frac{1}{N} \sum_{i,j=1}^N \bk{w_iw_j} \bk{\Phi_i(X_1)\Phi_j(X_2)} =
    \frac{1}{N} \sum_{i,j=1}^N\delta_{ij} 
\bk{\Phi_i(X_1)\Phi_j(X_2)} =  \bk{\Phi_i(X_1)\Phi_i(X_2)}
\end{equation}
with no sum on the $i$ on the RHS. We see the effect of the normalization: $G^{(2)}(X_1,X_2)$ does not scale with $N$. Restricting to the PS, we have 
\begin{equation}
    G^{(2)}_\varphi(x_1,x_2) = \frac{1}{x_{12}^{2\Delta}}
\end{equation}
where $\Delta = \Delta_\Phi$. Canonical normalization follows from the assumption on the second moment $\bk{w^2}$, and therefore we also have $G^{(2)}_\varphi(x_1,x_2)=G^{(2)}_\Phi(x_1,x_2)$.

The four-point function is 
\begin{equation}
    \begin{split}
        G^{(4)}_\varphi(X_1,X_2,X_3,X_4) &= \frac{1}{N^2} \sum_{i,j,k,l=1}^N\bk{w_iw_jw_kw_l} \bk{\Phi_i(X_1)\Phi_j(X_2)\Phi_k(X_3)\Phi_l(X_4)}.
    \end{split}
\end{equation}
A short computation yields 
\begin{equation}
    G^{(4)}_\varphi(X_1,X_2,X_3,X_4)=\frac{\gamma^4}{N} G^{(4)}_\Phi(X_1,X_2,X_3,X_4) + \left(1-\frac{1}{N}\right)\left[G^{(2)}_\Phi(X_1,X_2)G^{(2)}_\Phi(X_3,X_4) + \text{perms}\right],
\end{equation}
and from this it is clear that the connected four-point function satisfies $G^{(4)}_{\varphi,c}\propto 1/N$. Restricting to the CFT on the PS, we have 
\begin{equation}
    G^{(4)}_\varphi(x_1,x_2,x_3,x_4) = \left(x_{12}^2x_{34}^2\right)^{-\Delta}\, g_\phi(u,v)
\end{equation}
where
\begin{equation}\label{eq:g_mixing}
 g_\varphi(u,v) = \frac{\gamma^4}{N} g_\Phi(u,v) + \left(1-\frac{1}{N}\right)\left(1+u^\Delta + \left(\frac{u}{v}\right)^\Delta\right).
\end{equation}
We emphasize that this result is exact in $1/N$. We have two limiting cases of interest
\begin{align}
    N=1 &: \qquad g_\varphi(u,v) = \gamma^4 g_\Phi(u,v)\\
    N=\infty &: \qquad g_\varphi(u,v) = 1+u^\Delta + \left(\frac{u}{v}\right)^\Delta,
\end{align}
which shows us that the theory is interpolating between a simple rescaling of $\Phi$ at $N=1$ and a generalized free CFT of dimension $\Delta$ at $N=\infty$. This is how the NNGP correspondence arises in our CFT construction.

After doing some important additional work related to theories satisfying the unitarity bound, we will use this analysis to construct the free boson in Section \ref{sec:freeboson}.

\subsection{Deep Neural Networks and Recursive Conformal Fields}

Though our formalism applies to any Lorentz-invariant ensemble of homogeneous neural networks (of any such architecture), one might complain that our examples thus far, $\Phi_\Theta(X) = (\Theta\cdot X)^{-\Delta}$, are too simple to earn the title ``neural network." Recent progress in machine learning are driven by \emph{deep} neural networks of various architectures, amounting to the composition of many simpler parameterized architectures.

In this Section we derive a number of results about how our formalism manifests itself in the case of deep neural networks. The central results are:
\begin{itemize}
\item \textbf{CFT Layer.} A simple generalization of the above to $(\Theta_i \cdot X)^{-\Delta}$ defines a conformal input layer to the neural network, in the sense that any homogeneous network appended to it defines a conformal field (assuming finite correlators).
\item \textbf{Composition and Recursion}. Deep networks involve composition of many layers, and by appending it to a CFT layer we obtain a conformal field at each layer, the correlation functions of which depend recursively on the data of the previous layer.
\item \textbf{Deep Linear Networks.} The simplest case is a deep linear network, which preserves the conformal dimension at each layer. We give explicit recursion relations for the two-point and four-point correlators, as well as $g(u,v)$.
\end{itemize}           
We anticipate thorough exploration of these concepts in future work, but here provide the essentials since they are simple to understand and derive.

\vspace{1cm}
Let us begin with the idea of a CFT layer, which extends the construction to many more examples. Our first example $\Phi_\Theta(X)=(\Theta \cdot X)^{-\Delta}$ is a map from $\bR^{D+1,1} \to \bR$, which we may trivially extend to $\bR^N$ by growing in indices and parameters
\begin{equation}
\Phi^{(0)}: \bR^{D+1,1} \to \bR^N \qquad \qquad \Phi^{(0)}_i(X) = (\Theta_i \cdot X)^{-\Delta_{\Phi^{(0)}}},
\end{equation}
where we have omitted the $\Theta$ subscript for notational simplicity, introduced a superscript $(0)$ on $\Phi$ since we will think of this as the zeroth layer in a deep network, and have parameters $\Theta_{ij}$ an $N\times (D+2)$ matrix. $\Phi_i(X)$ is a conformal field provided that $P(\Theta_{ij})$ is Lorentz-invariant and the correlators are finite. Then the deeper network
\begin{equation}
g_{\Theta_g} \circ \Phi_i^{(0)}(X) \qquad \Theta_{g} \cap \Theta = \emptyset
\end{equation}
obtained by composition with homogeneous $g_{\Theta_g}$ with new parameters $\Theta_g$ also defines a conformal field (or fields, depending on the index structure of $g$) provided that the correlators are finite, since Lorentz-invariance in $(D+2)$-dimensions is ensured by the input layer $\Phi_i$ and the whole network is still homogeneous. By appending a homogeneous layer with new parameters to the CFT input layer, we obtain new conformal fields.

Deep neural networks are directly treatable with this argument. Consider a specific $g$, such that our deep network is of the form
\begin{equation}
\Phi^{(\ell)}(X) = f^{(\ell)}_{\Theta^{(\ell)}} \circ \dots \circ f^{(1)}_{\Theta^{(1)}}(\Phi_i(X)) \in \bR^{N_\ell}
\end{equation}
where $i=1,\dots,N_0$ and $f^{(j)}_{\Theta^{(j)}}:\bR^{N_{j-1}} \to \bR^{N_j}$ is the $j^{\text{th}}$ homogeneous layer with parameters $\Theta^{(j)}$ not appearing in any other layer. Of course, one may alternatively write this in terms of the network up to the previous layer as  
\begin{equation}
\Phi^{(\ell)}(X) =  f^{(\ell)}_{\Theta^{(\ell)}}(\Phi^{(\ell-1)}(X))
\end{equation}
For any $j$, $\Phi^{(j)}$ is homogeneous and its correlators are Lorentz-invariant, and thus yields a conformal field by restricting to the projective null cone. The conformal dimension depends on $\Delta_{\Phi^{(0)}}$ and the dimensions $\Delta_{f^{(j)}}$ of $f^{(j)}_{\Theta^{(j)}}$ as 
\begin{equation}
\Delta_{\Phi^{(j)}} = \Delta_{\Phi^{(0)}} \prod_{i=1}^{j} \Delta_{f^{(j)}},
\end{equation}
which satisfies the recursion relation
\begin{equation}
\Delta_{\Phi^{(j)}} = \Delta_{f^{(j)}} \Delta_{\Phi^{(j-1)}}.
\end{equation}
Since the conformal dimensions satisfy a recursion relation the two-point function is fixed, and the four-point functions depend on the data of the previous layer as 
\begin{equation}
G^{(4)}_{\Phi^{(\ell)}}(X_1,\dots,X_4) = \bk{f^{(\ell)}_{\Theta^{(\ell)}}(\Phi^{(\ell-1)}(X_1)) \dots f^{(\ell)}_{\Theta^{(\ell)}}(\Phi^{(\ell-1)}(X_4))}.
\end{equation}
At each successive layer we have a conformal field
\begin{equation}
\phi^{(\ell)}(x) = \Phi^{(\ell)}|_\text{PS}
\end{equation} 
that depends recursively on the data of the previous layer, with conformal correlators obtained by restriction to the Poincar\'e section.

Let us specialize further to one final case: a deep linear network of depth $L$ appended to a CFT layer, which generates a conformal field at each successive layer. We have 
\begin{align}
    \Phi^{(0)}_i(X) &= (\Theta_i \cdot X)^{-\Delta_{\Phi^{(0)}}} \qquad \Theta \sim P(\Theta) \\
    \Phi^{(\ell)}_i(X) &= W^{(\ell)}_{ij} \Phi^{(\ell-1)}_j(X) \qquad W^{(\ell)} \sim P(W^{(\ell)}) \qquad \ell = 1,\dots,L
\end{align}
with Einstein summation implied. The two-point and four-point functions satisfy
\begin{align}
 G^{(2), (\ell)}_{i,k} (X,Y)&= \bk{W_{ij}^{(\ell)}W_{kl}^{(\ell)}} \,\, G^{(2), (\ell-1)}_{j,l} (X,Y) \\
    G^{(4), (\ell)}_{ijkl} (X_1,\dots,X_4)&= \bk{W_{im}^{(\ell)}W_{jn}^{(\ell)}W_{ko}^{(\ell)}W_{lp}^{(\ell)}} \,\, G^{(4), (\ell-1)}_{m,n,o,p} (X_1,\dots,X_4),
\end{align}
recursion relations that relate the essential CFT data of the previous layer to the current layer. Again the conformal field and four-point function is obtained by restriction. In the $s$-channel decomposition this yields 
\begin{equation}
    g^{(\ell)}_{ijkl}(u,v) = \bk{W_{im}^{(\ell)}W_{jn}^{(\ell)}W_{ko}^{(\ell)}W_{lp}^{(\ell)}} \,\, g^{(\ell-1)}_{m,n,o,p}(u,v),
\end{equation}
which depends crucially on how the moment tensor on the RHS contracts with the $g(u,v)$ of the previous layer. This freedom allows from some degree of CFT engineering, governed by the random matrix theory associated to $P(W^{(\ell)})$ that is worthy of further study.

\section{Other Treatments of the Lorentzian Theory\label{sec:other_treatments}}

As discussed, our approach involves constructing a Lorentz-invariant theory on the embedding space and restricting it to a conformal field on the Poincar\'e section of the projective null cone. There are three essential ingredients: homogeneous and Lorentz invariance, which are relatively easy to ensure by appropriate choice of architecture and $P(\Theta)$, and the finiteness of the correlators, which is more difficult to ensure. 

In this section we will discuss another treatment of the Lorentzian theory that can lead to finite correlators, and also some pitfalls that arise in seemingly natural approaches.

\subsection{Natural Pitfalls}

Our hope in this section is that statement of some simple pitfalls might be illustrative to the reader, or inspire future work.
Specifically, our non-unitary example was solved by working with a rotationally invariant $(D+2)$-dimensional theory, solving for the correlators, Wick-rotating to Lorentzian $D+2$, and pushing down to conformal correlators on the PS. It is therefore natural to hope that various other approaches involving the Wick rotation of $(D+2)$-dimensional Euclidean objects might be useful. In our experience, it is unfortunately more difficult than expected. 

Consider first any rotationally invariant NN-FT in $(D+2)$-dimensional Euclidean space, where we assume rotational invariance is ensured via the mechanism of \cite{maiti2021symmetryviaduality} by choosing $P_E(\Theta)$ to be rotationally invariant. Now instead of evaluating the correlators and Wick rotating, as in Section \ref{sec:NonUnitary}, instead consider obtaining a Lorentz invariant $P(\Theta)$ from a Wick rotation of $P_E(\Theta)$. For instance, if the rotationally invariant $P_E(\Theta)$ is a multivariate Gaussian
\begin{equation}
P_E(\Theta) \propto \exp{\left(-\frac12 \frac{\Theta_0^2 + \vec \Theta \cdot \vec \Theta}{\sigma^2}\right)},
\end{equation}
after Wick rotation we have
\begin{equation}
    P(\Theta) \propto \exp{\left(-\frac12 \frac{-\Theta_0^2 + \vec \Theta \cdot \vec \Theta}{\sigma^2}\right)},
\end{equation}
which is Lorentz-invariant by construction but is no longer a probability distribution since it is not integrable. One might try to put in a Lorentz-invariant cutoff that avoids the singularity, but we have found no such simple solution. 

Another approach that one might take is to study Euclidean $(D+2)$-dimensional numerics, as in lattice field theory, in order to obtain information about the Lorentzian theory from the Euclidean correlators. We discuss such numerical approaches in Section \ref{sec:numerics}, but an essential difficulty is that it is hard to take a Euclidean numerical result to Lorentzian signature, unless one finds a good symbolic approximation to the Euclidean correlator; we have not attempted the latter, and think it is an interesting direction for future work.

Finally, another question is whether we might relax $P(\Theta)$ so that it need not be a probability density, but rather a function that one integrates operators against to obtain a notion of correlator. We are used to this distinction in the ordinary Euclidean and Lorentzian path integrals, 
\begin{equation}
\int D\phi \, e^{-S_E[\phi]} \qquad \text{vs.} \qquad \int D\phi\,  e^{iS[\phi]},
\end{equation}
where the Euclidean path integral has a straightforward probabilistic interpretation but the $e^{iS}$ of the Lorentzian path integral is more subtle. We remain open-minded about the possibilities in our context, as in the example of section \ref{sec:Unitary} we will eventually take
\begin{equation}
P(\Theta) = \frac{1}{\Theta^2+1},
\end{equation}
which is not a probability density because $P$ is negative when the Minkowki product $\Theta^2 < -1$. However, this $P$ is integrable in dimensional regularization and we will still be able to compute conformal correlators. Alternatively, a cutoff on $\Theta^2$ such that $\Theta^2 > -1$ yields a probability density, and one may take a straightforward numerical approach.

\subsection{Amplitudes Techniques}\label{sec:Unitary}

For our construction to work, we need well-behaved correlators, which are Lorentz-invariant integrals. The amplitudes community has developed many techniques for studying Feynman integrals, which are of this type, and in this section we review and apply their techniques in the context of our construction of conformal fields. Full details are in Appendix \ref{app:IBP_DE}. Here we highlight the setup, key conceptual insights, and the main results. 

The example at hand is similar to the non-unitary theory of Section~\ref{sec:NonUnitary}, but we now flip the sign on the scaling dimension in order to satisfy the unitarity bound in 4D. Our network is defined by 
\begin{equation}
    \Phi(X) = (\Theta\cdot X)^{-1} 
\end{equation}
and the inner product is taken in $\mathbb{R}^{D+1,1}$ with Minkowski metric. Via the dictionary provided by the embedding formalism between the objects in $\mathbb{R}^{D+1,1}$ and in $\mathbb{R}^{D}$, $\Phi(X)$ descends to a field $\phi(x)$ defined on the PS. 
Plugging $\Phi(X)$ into~(\ref{eq:general_partition_function}), the partition function is:
\begin{equation}\label{eq:unitary_partition_function}
    Z[J] = \int D\Theta P(\Theta) e^{\int d^{D+2}X \frac{J(X)}{\Theta\cdot X}}.
\end{equation}
We call this theory \ibp, and fully specifying it requires specifying $P(\Theta)$, which we will do in a moment.
To solve theory \ibp we need to evaluate the associated correlation functions 
\begin{equation}\label{eq:unitary_npt}
    G^{(n)}(X_1,\cdots,X_n) = \int D\Theta\,P(\Theta)\,\, \prod_{i=1}^n \frac{1}{\Theta\cdot X_i}.
\end{equation}
In order to have manifestly $SO(D+1,1)$ invariant $G^{(n)}$ \emph{at least formally}, we require $P(\Theta)$ to be Lorentz invariant, which we impose by having it depend on $\Theta$ through $\Theta^2$.

The construction of \ibp via~(\ref{eq:unitary_partition_function}) and~(\ref{eq:unitary_npt}) is rather different from the construction of non-unitary theories via~(\ref{eq:nonunitary_partition_function}) with field~(\ref{eq:nonunitary_field}) in two aspects. One aspect is very straightforward that rather than working with Euclidean metric, we now directly work with Minkowski metric which is the reason we drop the subscript $E$ in $Z[J]$ and $\Phi(X)$ compare to~(\ref{eq:nonunitary_partition_function}) and~(\ref{eq:nonunitary_field}). This means in principle that we may not have a probabilistic interpretation of $P(\Theta)$, since we don't have such an interpretation in normal Lorentzian field theory in general. However, working with Minkowski metric allows us to employ powerful amplitude techniques to calculate $G^{(n)}$, as we shall see in a moment.

The other aspect is more subtle. Unlike the non-unitary case, the appearance of $\Theta\cdot X$ in the denominator of the integrand of $G^{(n)}$ inevitably makes $G^{(n)}$ divergent for general $n$. Assuming the regularity of $P(\Theta)$ at $\Theta = 0$, this can be seen by looking at the ``IR limit'' $|\Theta|\rightarrow 0$, in which limit the integral~(\ref{eq:unitary_npt}) becomes:
\begin{equation}
    G^{(n)} \sim \int_0 \frac{|\Theta|^{D+1} d|\Theta|}{|\Theta|^n}
\end{equation}
which diverges when $n \geq D+2$. Similarly, if $P(\Theta)$ does not decay sufficiently fast in the ``UV limit'' $|\Theta| \rightarrow \infty$, $G^{(n)}$ will suffer UV divergence if $|\Theta|^n$ in the denominator fails to beat $P(\Theta)|\Theta|^{D+1}$ in the numerator in the UV limit. Therefore, in order to make sense of $G^{(n)}$, one has to regularize the naively divergent integral~(\ref{eq:unitary_npt}). 

To compute $G^{(n)}$, we note that the integral~(\ref{eq:unitary_npt}) takes exactly the same form of a 1-loop Feynman integral:
\begin{equation}
    J^{(s)}(\mathbf{n}) = J^{(s)}(n_0,\cdots,n_s):= \int d^d\Theta\, j^{(s)}(n_0,\cdots,n_s) = \int d^d\Theta \frac{1}{D_0^{n_0} \prod_{i=1}^s D_i^{n_i}},
\end{equation}
where $\Theta$ plays the role of the loop momentum and each of the $s$ $X_i$'s plays the role of an external momentum.
Here we have also defined $j^{(s)}$ as the integrand of the integral expression of $J^{(s)}$. It is easy to see that for theory \ibp we have:
\begin{equation}\label{eq:G=J}
    G^{(s)}(X_1,\dots,X_s) = J^{(s)}(1,\cdots,1)
\end{equation}
with:
\begin{equation}
    D_0 = 1/P(\Theta),\ D_i = \Theta\cdot X_i,
\end{equation}
where $d = D+2$. To treat conformal fields in our construction with amplitudes techniques, $P(\Theta)$ should be the sort of object that appears in a Feynman integral. For simplicity we take 
\begin{equation}
P(\Theta) = \frac{1}{\Theta^2 + 1},
\end{equation}
which looks like a propagator with $m^2=1$. Now the theory \ibp is completely defined by our choice of $P(\Theta)$ and $\Phi(X)$.

Due to the observation~(\ref{eq:G=J}), calculating the $s$-point function of \ibp is equivalent to evaluating the ``1-loop integral'' $J^{(s)}$, for which a detailed step-by-step calculation of $G^{(1)-(4)}$ is provided in Appendix~\ref{app:IBP_DE}. 
Since the result is an $SO(D+1,1)$-invariant scalar, we expect $J^{(s)}$ be a function of $X_{ij} := X_i\cdot X_j$, $1\leq i,j\leq s$ if the integration over $\Theta$ can be carried out. We will see in the following derivations that this requirement indeed leads to self-consistent meaningful results. We also note that in the embedding formalism in order to obtain a CFT in $\mathbb{R}^D$, \ibp shall be restricted to the projective null-cone of $\mathbb{R}^{D+1,1}$. In going from \ibp to the corresponding CFT we should impose $X_{ii} = 0$ to $J^{(s)}$. In other words, for the CFT $s$-point function we will have $J^{(s)}|_{X_{11} = \cdots X_{ss} = 0}$.

To evaluate $J^{(s)}$, we apply the Integration By Parts (IBP) identities~\cite{Chetyrkin:1981qh, Tkachov:1981wb}. The key fact leading to IBP identities is the following equation in dimensional regularization:
\begin{equation}\label{eq:IBP_J}
    \mathcal{D}_q J^{(s)}(n_0,n_1,\cdots,n_s) = 0,\ \forall q \in \{ \Theta,X_1,\cdots,X_s \}
\end{equation}
where the differential operator $\mathcal{D}_q$ is defined as:
\begin{equation}\label{eq:IBP_F}
    \mathcal{D}_q F := \int d^d\Theta \frac{\partial}{\partial \Theta} \cdot (qf)
\end{equation}
for $F = \int d^d\Theta f$. The validity of~(\ref{eq:IBP_J}) comes from the fact that $j^{(s)} \rightarrow 0$ as $|\Theta|\rightarrow \infty$. 
The identity~(\ref{eq:IBP_J}) leads to recursion relations that relates $J^{(s)}(\mathbf{n})$ with $J^{(s)}(\mathbf{m})$ for $n_0\geq m_0,\ \cdots,\ n_s\geq m_s$. As a concrete example, let us consider $J^{(1)}(n_0,n_1)$ which corresponds to 1-point function $G^{(1)}$ of \ibp when $n_0 = n_1 = 1$. Applying~(\ref{eq:IBP_J}) to $J^{(1)}(n_0, n_1)$ leads to the recursion relation:
\begin{equation}
    J^{(1)}(n_0+1,n_1) = \frac{2n_0+n_1-d}{2n_0} J^{(1)}(n_0,n_1).
\end{equation}
By the above recursion relation, the calculation of any $J^{(1)}(n_0,n_1)$ boils down to calculating $J^{(1)}(1,n_1)$, which is called \emph{master integral} in the literature which can be roughly viewed as the terminating point of the recursion relation obtained from the identity~(\ref{eq:IBP_J}). Many other IBP identities are derived in appendix \ref{app:IBP_DE}.

Some recursion relations are directly related to essential results in CFT. For instance, the recursion relation for $J^{(2)}(n_0,n_1,n_2)$ from~(\ref{eq:IBP_J}) is:
\begin{equation}
    \begin{split}
        &2n_0 J^{(2)}(n_0+1,n_1,n_2) \\
        =&\  (2n_0 + n_1 + n_2 - d) J^{(2)}(n_0,n_1,n_2) \\
        =&\  -n_2(X_1\cdot X_2) J^{(2)}(n_0, n_1+1, n_2+1) \\
        =&\  -n_1(X_1\cdot X_2) J^{(2)}(n_0, n_1+1, n_2+1)
    \end{split}
\end{equation}
The last equality of the above recursion relation, which is of particular interest to us, implies:
\begin{equation}\label{eq:vanishing_G2_neq}
    n_1 J^{(2)}(n_0,n_1,n_2) = n_2 J^{(2)}(n_0,n_1,n_2)
\end{equation}
which, when $n_1\neq n_2$, is consistent only if $J(n_0, n_1, n_2) = 0$. A direct consequence of this conclusion is $J^{(2)}(1, n_1, n_2) = 0$ for $n_1 \neq n_2$, which becomes:
\begin{equation}\label{eq:vanishing_G2}
    \int d\Theta P(\Theta) (\Theta\cdot X_1)^{-n_1}(\Theta\cdot X_2)^{-n_2} = 0,\ \text{for},\ n_1\neq n_2.
\end{equation}
The above result, after restricting to PS, can immediately be reinterpreted as the desired fact that a CFT 2-point function between fields with different scaling dimensions vanishes.

After making use of the recursion relation to reduce the calculation of general $J^{(E)}(\mathbf{n})$ to the calculation of a set of master integrals, we still must still evaluate the master integrals. To do this we employ the differential equation method~\cite{Gehrmann:1999as, Gehrmann:2000zt, Gehrmann:2001ck, Kotikov:1990kg, Kotikov:1991hm, Kotikov:1991pm}. The master integrals obey a set of PDEs (see Appendix~\ref{app:IBP_DE} for the derivation):
\begin{equation}\label{eq:chain_rule}
    \frac{\partial f}{\partial X_{ij}} = \sum_{n=1}^{D+2} \frac{\partial f}{\partial X_{in}} \sum_{k=1}^s X_{kn} \mathbb{G}^{-1}_{kj} = \sum_{k=1}^s \mathbb{G}^{-1}_{kj} X_k \cdot \frac{\partial f}{\partial X_i}
\end{equation}
for a master integral $f$ of all $X_{ij}$'s where $X_{in}$ is the $n^{\text{th}}$ component of $X_i$ and $\mathbb{G}_{ij} = X_{ij}$.

\bigskip
With this backdrop, we finally turn to the correlation functions. The odd-point functions vanish by symmetry, but we must compute the two-point and four-point function.

We now calculate the 2-point function $G^{(2)} \equiv J^{(2)}(1,1,1)$. Applying~(\ref{eq:chain_rule}) with $(i,j)=(1,1)$ to $J^{(2)}(1,n,n)$ we have:
\begin{equation}
    \frac{\partial J^{(2)}(1,n,n)}{\partial X_{11}} = -\frac{n}{X_{11}} J^{(2)}(1,n+1,n-1) = 0
\end{equation}
due to~(\ref{eq:vanishing_G2_neq}). It is easy to show that $\frac{\partial J^{(2)}(1,n,n)}{\partial X_{22}} = 0$ in the completely same manner. Thus we conclude that $J^{(2)}(1,n,n)$ depends only on $X_{12}$. Now applying~(\ref{eq:chain_rule}) with $(i,j) = (1,2)$ to $J^{(2)}(1,n,n)$ we have:
\begin{equation}
    \frac{\partial J^{(2)}(1,n,n)}{\partial X_{12}} = - \frac{n}{X_{12}} J^{(2)}(1,n,n).
\end{equation}
The solution to the above DE is:
\begin{equation}
    J^{(2)}(1,n,n) = \frac{C}{X_{12}^n}
\end{equation}
with integral constant $C$. Hence when $n=1$ we have:
\begin{equation}\label{eq:unitary_2pt}
    G^{(2)}(X_1,X_2) = \frac{C}{X_1\cdot X_2}
\end{equation}
A further restriction of $G^{(2)}(X_1,X_2)$ to the PS leads to
\begin{equation}
    G^{(2)}(x_1,x_2) = \frac{C}{(x_1-x_2)^2}.
\end{equation}
We see that is has  the structure required by conformal invariance.

A similar calculation using the IBP identities and differential equations fixes the four-point function up to an overall constant, see Appendix \ref{app:IBP_DE}. It is given by
\begin{equation}\label{eq:unitary_4pt}
    \begin{split}
        &G^{(4)}(X_1,X_2,X_3,X_4) \\
        =&\ \frac{C}{\sqrt{X_{12}^2 X_{34}^2 + X_{13}^2 X_{24}^2 + X_{14}^2 X_{23}^2 - 2X_{12} X_{34} X_{13} X_{24} -  2X_{12} X_{34} X_{14} X_{23} - 2 X_{13} X_{24} X_{14} X_{23}}},
    \end{split}
\end{equation}
where we have already used the $X_i^2=0, i=1,\dots,4$, condition associated to the null cone. Pushing to the Poincar\'e section and using $s$-channel variables we have 
\begin{equation}
    G^{(4)}(x_1,x_2,x_3,x_4) = \frac{C}{x_{12}^2 x_{34}^2} \frac{u}{\sqrt{u^2 + v^2 - 2uv - 2u - 2v + 1}}
\end{equation}
which matches the general form of $G^{(4)}$ given by~(\ref{eq:4ptallphi}) with $\Delta = 1$ and:
\begin{equation}
    g(u,v) = \frac{C u}{\sqrt{u^2 + v^2 - 2uv - 2u - 2v + 1}}.
\end{equation}
This is conformally invariant and satisfies the crossing conditions \eqref{eq:crossing}.
Let $u = z\Bar{z}$ and $v = (1-z)(1-\Bar{z})$ as in~(\ref{eq:zx-variables}) and set $C = 1$ for simplicity, we have:
\begin{equation}
    g(u,v) \rightarrow g(z,\Bar{z}) = \frac{z\Bar{z}}{|\Bar{z} - z|},
\end{equation}
but this cannot be expanded at $z=\Bar{z}$ as is usually done in the CBD~\cite{Rattazzi:2008pe}. 

We think that this uncommon feature is an accident of this example, and we are not sure of its interpretation. It is against standard CFT expectations, and yet we have a theory of fields with conformally invariant correlators. We hope to understand when and why such features arise by studying more examples in the future.

\subsection{Numerical Approaches \label{sec:numerics}}
It would be neither modern CFT nor machine learning if we said nothing about a numerical approach. In all cases, our correlators and associated CFT data are represented by Lorentz-invariant correlators 
\begin{equation}
G^{(n)}(X_1,\dots,X_n) = \int D\Theta \,\, P(\Theta) \Phi(X_1) \dots \Phi(X_n),
\end{equation}
that yield conformal correlators by restriction.
Various methods of numerically evaluating the integral allow in principle for the extraction of CFT data. We do not carry out these methods here, as our focus is on the general construction and also because simple approaches are more difficult than one might naively expect.

Nevertheless, we would like to state some of the possibilities in case the reader is interested in pursuing a numerical approach:
\begin{enumerate}
\item \textbf{Monte Carlo Integration.} The simplest approach is to sample $\Theta$ from $P(\Theta)$ $M$ times and evaluate the correlators
\begin{equation}
G^{(n)}(X_1,\dots,X_n) \simeq \frac{1}{M} \sum_{\Theta \sim P(\Theta)} \Phi(X_1) \dots \Phi(X_n).
\end{equation} 
Monte Carlo sampling has been used to compute neural network correlators in other works, e.g. \cite{Halverson_2021}. Notably, it is easy to write a Lorentz-invariant $P(\Theta)$, but difficult to ensure that it is a normalizable probability distribution.

\medskip
Alternatively, if $P(\Theta)$ is normalizable but takes negative values for some $\Theta$, then one might instead restrict the domain to values of $\Theta$ where $P(\Theta) > 0$ and sample from this probability density using Monte Carlo techniques. For instance, in our example where $P(\Theta) = 1/(\Theta^2+1)$, one might impose a Lorentz invariant IR cutoff such that $\Theta^2 > -1+\epsilon$, sample, and study the associated Lorentzian theory. 

\item \textbf{Monte Carlo Integration of Euclidean Correlators.} Alternatively we may begin with a Euclidean case on $\bR^{D+2}$ as in Section \ref{sec:NonUnitary}. The density $P(\Theta)$ is rotationally invariant to ensure rotationally invariant Euclidean correlators, which may be evaluated with Monte Carlo integration by drawing $\Theta$ from $P(\Theta)$. This is straightforward, as simple $P(\Theta)$ such as the multivariate i.i.d. Gaussian are rotationally invariant, proper probability densities, and easy to sample. A difficulty is that standard NN computations of these correlators are on a set of specific $X$-values, making analytic continuation difficult. This could be circumvented by symbolic regression of the correlators, such as with a Kolmogorov-Arnold network \cite{Liu:2024swq}.
\item \textbf{Direct Integration.} Alternatively, one can try to do the numerical integration directly, which might suffice for simple neural networks with few enough parameters. 
\end{enumerate}

\section{The Free Boson and Mixing
\label{sec:freeboson}}

In this Section we would like to take some of the techniques that we have developed and apply them to obtain the standard free boson as an example. We will show that we may combine the free boson and interacting conformal fields by stacking the conformal field or mixing them together by addition. We apply both to Theory \ibp, demonstrating how the techniques can add identity contributions to the four-point function.

Crucially, these stacking and adding techniques make the (potentially rescaled) free boson four-point function appear in the interacting four-point function, ensuring the existence of a stress tensor due to the contribution of a dimension $D$ spin-2 field to the CBD.

\subsection{The Free Boson as a NN-CFT}

In this Section we construct the free boson. We build on the results of Section \ref{sec:freetheory}, where we showed how to construct a general free conformal field $\varphi$ of dimension $\Delta$ out of $N\to\infty$ neurons $\Phi$ of the same dimension upon restriction to the Poincar\'e section of the PNC. The construction assumed only that the neuron correlators are themselves well-behaved, which allowed for the use of the CLT to eliminate non-Gaussianities. Since the free boson is unitary, we needed a construction of theories satisfying the unitarity bound before proceeding, which we obtained in Section \ref{sec:other_treatments} for \ibp.

Recalling the essentials from Section \ref{sec:freetheory}, the field 
\begin{equation}
\varphi(X) = \frac{1}{\sqrt{N}} \sum_i w_i \Phi_i(X) \qquad \qquad X\in \bR^{D+1,1},
\end{equation}
yields a generalized free CFT of dimension $\Delta$ provided that $w_i$ and $\Phi_i$ are each i.i.d. and that the correlators of $\Phi_i$ are themselves conformally invariant upon restriction to the PNC.
Therefore, any neuron satisfying 
\begin{equation}
\Delta_\Phi = \frac{D-2}{2}
\end{equation}
has an associated $\varphi$ that yields a free boson on the PNC in the $N\to \infty$ limit. We emphasize the generality of the construction with many possible realizations.

\bigskip
As a concrete example of the general construction let each $\Phi_i$ be a copy of the theory \ibp from Section~\ref{sec:Unitary}, which satisfies the unitarity bound. Using techniques from the amplitudes community we computed the exact two-point~(\ref{eq:unitary_2pt}) and four-point~(\ref{eq:unitary_4pt}) function of the conformal field $\Phi$. Putting indices on that example, we have
\begin{equation}
\Phi_i(X) = \frac{1}{\Theta_i\cdot X} \qquad \qquad P(\Theta_i) = \frac{1}{\Theta_i^2 + 1},
\end{equation}
where $\Theta_i^2$ uses the $(D+2)$-dimensional Minkowski metric.  Matching the example to the free boson scaling dimension, we deduce that $D=4$. A technical point is that the zero-point function  $\int D\Theta \, P(\Theta)$ diverges in $D=4$, but is finite via dimesional regularization in $D=4-\epsilon$ dimensions; we henceforth ignore this subtlety, computing in $4-\epsilon$. Since this is the free boson CFT in $D=4$, the rest of its properties follow, but we would like to see some of them emerge in the NN context.

Specifically, stress tensor is
\begin{equation}\label{eq:TT_D}
 T_{\mu\nu} = \varphi \partial_i \partial_j \varphi - 2\left[\partial_i \varphi \partial_j \varphi - \frac14 \delta_{ij} (\partial \varphi)^2 \right]
\end{equation}
We may compute its two-point function in the parameter space description and write it as (after descending down to $D$-dimensions), 
\begin{equation}
\bk{T_{\mu\nu}(x)T_{\rho\sigma}(y)}=\frac{c}{\mathcal{N}}\frac{1}{(x-y)^8} \left( \frac{1}{2}\left( I_{\mu\rho}(x-y)I_{\nu\sigma}(x-y)+I_{\mu\sigma}(x-y)I_{\nu\rho}\right)(x-y) - \frac{1}{4} \delta_{\mu\nu}\delta_{\rho\sigma} \right)
\end{equation}
where $c$ is the central charge with respect to normalization $\mathcal{N}$ and $I_{\mu\nu}(z)$ is defined as
\begin{equation}
I_{\mu\nu}(z)=\delta_{\mu\nu}-2 \frac{z_{\mu}z_{\nu}}{z^2}.
\end{equation}
See Appendix \ref{app:stress-tensor} for details of the computation.

\subsection{Mixing, Identities, and Stress Tensors}

Central to neural networks is the composition of simpler functions, which in our CFT context motivates investigating how pieces can be put together to achieve various aims. Some of the theories we have studied are missing identity blocks and / or stress tensors, and in this section we would like to incorporate them by mixing in generalized free fields or (more specifically) the free boson. These modify the theory, but leave it interacting.

There are two types of ``mixing" that we will study, which we call stacking and adding. We studied the stacking explicitly in Section \ref{sec:freetheory}, where we stacked $N$ independent copies of a theory $\Phi$ and then summed them up according to 
\begin{equation}
\varphi(X) = \frac{1}{\sqrt{N}} \sum_{i=1}^N w_i \Phi_i(X) \qquad \qquad X\in \bR^{D+1,1},
\end{equation}
where the normalization $1/\sqrt{N}$ is for Gaussianity as $N\to \infty$ and $w \sim P(w)$ i.i.d. is chosen to have unit variance so that the two-point function is preserved. The cross-ratio dependent contribution to the four-point function (on the PS) is 
\begin{align}
    \label{eqn:guv_stacking}
    g_\varphi(u,v) &= \frac{\gamma^4}{N} g_\Phi(u,v) + \left(1-\frac{1}{N}\right)\left(1+u^\Delta + \left(\frac{u}{v}\right)^\Delta\right) \\ &= \frac{\gamma^4}{N} g_\Phi(u,v) + \left(1-\frac{1}{N}\right)\, g_{\text{free},\Delta}(u,v),
\end{align}
which interpolates between the $\Phi$ theory and the generalized free CFT of dimension $\Delta$ as $N\to \infty$. In this sense, the $\varphi$ theory takes the $\Phi$ theory and mixes in a generalized free contribution, with the degree of mixing set by $N$.

The other type of mixing is adding, where we take $\Phi(X)$ of dimension $\Delta$ and $\Gamma(x)$ a generalized free field of dimension $\Delta$ and add them together to get a new field 
\begin{equation}
\varphi(X) = \frac{1}{\sqrt{2}} \bigg(\Phi(X) + \Gamma(X)\bigg).
\end{equation}
The normalization preserves the two-point function $G^{(2)}_\varphi = G^{(2)}_\Phi = G^{(2)}_\Gamma$ and the four-pount function is 
\begin{equation}
G^{(4)}_\varphi(X_1,\dots,X_4) = G^{(4)}_\Phi(X_1,\dots,X_4) + G^{(4)}_\Gamma(X_1,\dots,X_4) + 2 \left[\frac{1}{X_{12}X_{34}} + \text{2 perms}\right],
\end{equation}
where the latter contribution is from products of $G^{(2)}_\Phi$ and $G^{(2)}_\Gamma$. Since they are the same, the quantity in square brackets also equals $G_{\Gamma}^{(4)}$, and we have 
\begin{equation}
G^{(4)}_\varphi(X_1,\dots,X_4) = G^{(4)}_\Phi(X_1,\dots,X_4) + 3 \,G^{(4)}_\Gamma(X_1,\dots,X_4).
\end{equation}
The cross-ratio dependent contribution to the four-point function (on the PS) is therefore 
\begin{equation}
    \label{eqn:guv_adding}
g_\varphi(u,v) = g_\Phi(u,v) + 3\, g_{\text{free},\Delta}(u,v).
\end{equation}
where we have used the notation $g_{\text{free},\Delta} = g_\Gamma$ for easy comparison to \eqref{eqn:guv_stacking}.

In summary, stacking and adding both modify the theory in a way that preserves the two-point function, but mixes generalized free contributions into the four-point function and generally leaves the theory interacting. As immediate corollaries: 
\begin{itemize}
\item \textbf{Identity.} Since the generalized free CFT has an identity block, both stacking and adding will introduce identity contributions to the four-point function.
\item \textbf{Stress Tensor.} When $\Delta = (D-2)/2$ the generalized free theory is the free boson, and both the stacking and adding theory inherit its stress tensor.
\end{itemize}
In particular: since theory \ibp has neither an identity nor a stress tensor contribution to $g_{\Phi}(u,v)$, they can be introduced by either way of mixing in the free boson.

\section{Discussion and Outlook\label{sec:discussion}}

In this paper we present a novel construction of conformal fields that leverages the embedding formalism and neural networks. The construction relies on three crucial principles: 
\begin{itemize}
\item \textbf{Homogeneity} is achieved by specifying a homogeneous neural network architecture $\Phi_\Theta(X)$ on the embedding space $\bR^{D+1,1}$, with parameters $\Theta$.
\item \textbf{Lorentz invariance} arises via the choice of Lorentz-invariant $P(\Theta)$, interpreted as a probability distribution over the parameters of the neural network in the canonical context, though we allow for more general interpretation as well.
\item \textbf{Finite correlators.} The correlators should not be divergent everywhere.
\end{itemize}
While homogeneous and Lorentz invariance are straightforward to achieve, ensuring that the correlators are finite takes some work, and we present a number of approaches to that problem. Once the three principles are satisfied, one has a Lorentz-invariant field on the embedding space (that is not necessarily translation invariant) that restricts to a conformal field on the Poincar\'e section of the projective null cone. Unlike previous studies using the embedding formalism, which focus on kinematical constraints, in our work we use the embedding formalism to construct conformal fields.

These are the essentials of the formalism. Given them, a number of other results follow naturally:
\begin{itemize}
\item \textbf{Free theory limits} arise in a large-$N$ limit by the Central Limit Theorem when summing i.i.d. neurons. This allows for the construction of generalized free fields and the free boson, and is known as the neural network / Gaussian process correspondence in the machine learning literature.
\item \textbf{Conformal layers} at input ensure conformal symmetry at each layer of the network, provided that each subsequent layer is homogeneous and has its own parameters. The conformal fields at successive layers are related by recursion relations on their conformal dimensions, and in some cases on their four-point functions.
\item \textbf{Numerical approaches} of different types, including Monte Carlo integration and direct integration, can in principle be used to evaluate the correlators of the conformal field. This is an important direction for future work.
\end{itemize}
We also study a number of examples, including an exactly solvable non-unitary theory, and a theory satisfying the unitarity bound in $4D$ that we solve with amplitudes techniques.

In the Introduction, we discussed ``what constitutes a CFT?" Our formalism guarantees and ensemble of field configurations with conformally invariant correlators, which we take as a minimal definition of a CFT. Nevertheless, in our non-unitary example a non-trivial calculation of OPE coefficients demonstrates the conformal block decomposition of the exact (self) four-point function agrees with the OPE coefficients in precisely the expected way. However, in the mixed correlator of the form $\langle \Phi \Phi \Phi^2 \Phi^2\rangle$ we saw that an additional constraint on the moments of $P(\Theta)$ must be satisfied in order to obtain the OPE / CBD matching. Understanding the details of these additional moment constraints and whether they can be satisfied is an important direction for future work.

Despite the mismatch in the mixed correlator, an important principle emerged in obtaining the OPE matching for the self-correlator  $\langle \Phi \Phi \Phi \Phi\rangle$: the OPE should be homogeneous in parameters. This was crucial to getting the match, but we also argued for its relevance on general grounds. Related, the OPE motivates the idea that the different operators in the theory should have the same parameter distributions. This stems from the fact that the OPE should hold as an operator equation, and therefore under any expectation value. If the different operators in the OPE have different parameter distributions, then the definition of the expectation changes from one operator to another, which would be rather unusual. This idea was reflected in the non-unitary example, where the operators that led to the OPE / CBD matching had the same parameter distribution.

One may easily deform a theory in our construction, which deserves some discussion. Let $(\Phi_\Theta, P(\Theta))$ be a pair that defines a conformal field upon restriction to the PS. One may deform the theory by either architecture deformations
\begin{equation}
(\Phi_\Theta,P(\Theta)) \mapsto (\Phi_\Theta + \delta \Phi_\Theta, P(\Theta) + \delta P(\Theta)),
\end{equation}
where for simplicity we have assumed that any architecture deformation $\delta\Phi_\Theta$ depends only on $\Theta$ and does not introduce new parameters. A fixed-architecture deformation still satisfies homogeneity, and is Lorentz invariant if $\delta P(\Theta)$ is Lorentz invariant. It is straightforward to write down such deformations $\delta P$, and then the remaining condition to satisfy is to ensure that the correlators are finite: if so, then $(\phi_\Theta, P(\Theta) + \delta P(\Theta))$ is a deformation that defines a conformal field on the Poincar\' e section; we have performed a marginal deformation.

Naively this seems too easy, as a large number of such deformations might exist, and yet in many cases, conformal fixed points are isolated and do not have marginal deformations. However, the deformations we have discussed need not be deformations by integrated local operators, and they need not preserve unitarity, which may not have existed even in the undeformed theory. On the other hand, when we consider isolated unitary CFTs, we usually mean that there are no marginal deformations by integrated local operators that preserve unitarity. This is a clear distinction from the notion that we have introduced, but it would be interesting to better understand the interplay in future work. 

A notable aspect of these deformations is that, by construction, they preserve the conformal invariance of the correlators, but in general they might not exhibit matching between the conformal block decomposition of multiple correlators and OPE coefficients. If this can be demonstrated, then the question of the Introduction is answered in the negative: conformal correlators are not enough to ensure the standard CFT assumptions utilized in the bootstrap. One might take this as a broader definition of a CFT, or perhaps use different language for field theories of this type that have conformal invariance. An analogous change of language is ``CTs," which are sometimes used in the literature \cite{Schwarz:2015fva} and lecture notes \cite{Kaplan2015} to describe CFTs that do not have stress tensors. 

On the other hand, we obtained the $D=4$ free boson by appropriately stacking-and-summing $N$ copies of a $\Delta=1$ NN-CFT (such as the theory \ibp that we studied in detail) and taking $N\to\infty$. Backing off the $N\to \infty$ limit corresponds to deforming the theory by a discrete architecture deformation that preserves the conformal dimension but introduces interactions. However, since the free boson is the unique $\Delta = 1$ unitary CFT in 4D \cite{Mack1977}, we deduce that the finite-$N$ realizations of our theory must break unitarity. This is perhaps not surprising, as nothing in our construction yet ensures reflection positive correlators, and a theory may have reflection positive $G^{(2)}$ but not have reflection positive higher correlators; e.g. this example has $G^{(2n)}_c\propto 1/N^{n-1}$, allowing for breaking reflection positivity at finite $N$ even if it is preserved as $N\to \infty$. See \cite{halverson2021building} for a discussion of reflection positivity.

\bigskip
There are a number of other interesting directions for future work. One is to undertake a numerical approach to approximate correlators, perhaps via one of the possibilities discussed in Section \ref{sec:numerics}. An interesting aspect of this is to couple it with a learning mechanism to drive it toward some desired CFT, by regressing on the hyperparameters of the architecture and $P(\Theta)$. One might also study the construction of spinning conformal fields, which have been studied in the embedding formalism. We see no obstruction to extending our results to that case: it requires more complicated architectures, but our principles of homogeneity, Lorentz invariance, and finiteness should still persist. Finally, one would like to have a better understanding of unitarity, by understanding how to specify $(\Phi_\Theta, P(\Theta))$ such that the conformal correlators are reflection positive on the Poincar\'e section. Reflection positivity is a difficult problem in NN-FT, for which a general answer is not yet known, but perhaps the difficulty could be alleviated by specifying to conformal theories.

\bigskip \noindent
\textbf{Acknowledgements.}
We thank Wei Fan, Sergei Gukov, Ziming Ji, Mario Martone, Alexander Migdal, Sridip Pal, Sylvain Ribault, Brandon Robinson, Laurentiu Rodina and Ning Su for discussions. We are especially grateful to Chi-Ming Chang, Liam Fitzpatrick, Sarah Harrison, Yikun Jiang, Costis Papageorgakis and Jiaxin Qiao for discussing details of the project and unusual aspects of CFTs, to Julio Parra-Martinez for introducing us to the package \texttt{LiteRed2}, to Jiaqi Chen for explaining certain subtle aspects of the amplitude methods, and to Christian Ferko for discussions and feedback on a draft. We would also like to thank the referee for thoroughly reading our work, and for extensive comments that suggested new calculations and pointed out a number of ways to improve the manuscript.
This research was supported in part by grant NSF PHY-2309135 to the Kavli Institute for Theoretical Physics (KITP).
This work is supported by the National Science Foundation
under Cooperative Agreement PHY-2019786 (The NSF
AI Institute for Artificial Intelligence and Fundamental
Interactions). J.H. is supported by NSF CAREER grant PHY-1848089. J.T. is supported by National Natural Science Foundation of China under Grant No. 12405085 and by the Natural Science Foundation of Shanghai (Grant No. 24ZR1419300). J.N. is partially supported by the NSF IAIFI and Northeastern University.

\appendix

\section{Non-locality in Quartic Vertex}
\label{app:non-local}

In this section we will show that the theory with 2-point and 4-point functions~(\ref{eq:delta-1_g2_g4}) is non-local by computing its quartic vertex applying the method in~\cite{Demirtas:2023fir}.

Our goal is to compute quartic vertex of the $\Delta = -1$ non-unitary theory using Eq.~(3.24) of~\cite{Demirtas:2023fir}:
\begin{equation}\label{eq:quartic_coupling}
	\begin{split}
		g_4(x_1,x_2,x_3,x_4) &= -\frac{1}{4!} \int dy_1dy_2dy_3dy_4\ G^{(4)}_c(x_1,x_2,x_3,x_4) \\
		&\qquad \times \left( G^{(2)}(y_1,x_1)^{-1} G^{(2)}(y_2,x_2)^{-1} G^{(2)}(y_3,x_3)^{-1} G^{(2)}(y_4,x_4)^{-1} + \text{perms} \right)
	\end{split}
\end{equation}
and for simplicity we ignore the permutations. For the field $\Phi(X) = \Theta\cdot X$, the rescaled $G^{(4)}$ is:
\begin{equation}
	\begin{split}
		G^{(4)} &= \frac{\mu_4}{3}(x_{12}^2x_{34}^2 + x_{13}^2x_{24}^2 + x_{14}^2x_{23}^2) = \frac{\mu_4}{3} (G^{(2)}(x_1,x_2)G^{(2)}(x_3,x_4) + \text{perms})
	\end{split}
\end{equation}
where at the second step we have used~(\ref{eq:delta-1_g2_g4}). Therefore we have:
\begin{equation}\label{eq:G4c_largeN}
	\begin{split}
		G^{(4)}_c(x_1,x_2,x_3,x_4) &= \frac{\gamma^4}{N}G_{\Phi}^{(4)}(x_1,x_2,x_3,x_4) + \left( 1 - \frac{1}{N}\right) (G^{(2)}(x_1,x_2)G^{(2)}(x_3,x_4) + \text{perms}) \\
		&\qquad - ((G^{(2)}(x_1,x_2)G^{(2)}(x_3,x_4) + \text{perms})) \\
		&= \frac{\gamma^4}{N}G_{\Phi}^{(4)}(x_1,x_2,x_3,x_4) - \frac{1}{N} (G^{(2)}(x_1,x_2)G^{(2)}(x_3,x_4) + \text{perms}) \\
		&= \frac{1}{N}\left( \frac{\gamma^4\mu_4}{3} - 1 \right) (G^{(2)}(x_1,x_2)G^{(2)}(x_3,x_4) + \text{perms})\,.
	\end{split}
\end{equation}
Applying the method in~\cite{Demirtas:2023fir}, we compute $g_4$ to be (here we display explicitly only the terms that are not in ``perms'' in~(\ref{eq:quartic_coupling}) and~(\ref{eq:G4c_largeN})):
\begin{equation}\label{eq:g4^1}
	\begin{split}
		g_4 &\propto \frac{1}{N}\left( \frac{\gamma^4\mu_4}{3} - 1 \right) \left( \int dy_1dy_2dy_3dy_4\ G^{(2)}(y_1,y_2) G^{(2)}(y_3,y_4) \right. \\
		&\qquad \left. \times G^{(2)}(y_1,x_1)^{-1} G^{(2)}(y_2,x_2)^{-1} G^{(2)}(y_3,x_3)^{-1} G^{(2)}(y_4,x_4)^{-1} + \cdots \right) \\
		&= \frac{1}{N}\left( \frac{\gamma^4\mu_4}{3} - 1 \right) \left( \int dy_2 dy_4\ \delta(x_1 - y_2) \delta(x_3 - y_4) G^{(2)}(y_2,x_2)^{-1} G^{(2)}(y_4,x_4)^{-1} + \cdots \right) \\
		&= \frac{1}{N}\left( \frac{\gamma^4\mu_4}{3} - 1 \right) \left( G^{(2)}(x_1,x_2)^{-1} G^{(2)}(x_3,x_4)^{-1} + \cdots \right)\,.
	\end{split}
\end{equation}

Even without detailed calculation one can see that $g_4$ is a non-local quartic coupling by reductio ad absurdum. Clearly, the following equation holds by definition of inverse operator:
\begin{equation}\label{eq:contradict_1}
	\begin{split}
	    \int dx_2dx_4dy\ G^{(2)}(x_1,x_2)^{-1} G^{(2)}(x_3,x_4)^{-1} \times G^{(2)}(x_2,y) G^{(2)}(x_4,y) &= \int dy \delta(x_1-y)\delta(x_3-y) \\
        &= \delta(x_1-x_3)\,.
	\end{split}
\end{equation}
While if $g_4$ were a local vertex then we must have $g_4 \propto \delta(x_1-x_2)\delta(x_1-x_3)\delta(x_1-x_4)$ for which we must have:
\begin{equation}\label{eq:contradict_2}
	\begin{split}
		&\int dx_2dx_4dy\ \delta(x_1-x_2)\delta(x_1-x_3)\delta(x_1-x_4) G^{(2)}(x_2,y) G^{(2)}(x_4,y) \\
        =&\ \int dy\ G^{(2)}(x_1,y) G^{(2)}(x_1,y) \delta(x_1-x_3) \\
		\neq&\ \delta(x_1-x_3)
	\end{split}
\end{equation}
since $\int dy\ G^{(2)}(x_1,y) G^{(2)}(x_1,y) \neq 1$. Since~(\ref{eq:contradict_1}) contradicts~(\ref{eq:contradict_2}), we conclude that:
\begin{equation}\label{eq:nonlocal_condition}
	g_4 \neq c \delta(x_1-x_2)\delta(x_1-x_3)\delta(x_1-x_4)
\end{equation} 
for any constant $c$. Hence, given~(\ref{eq:g4^1}) the theory must be non-local when $\frac{\gamma^4\mu_4}{3} \neq 1$.

\section{Concrete Example with Amplitudes Techniques}\label{app:IBP_DE}

We direct the readers to~\cite{Smirnov:2012gma, henn2015lectures} and the references therein for a thorough introduction to all the techniques applied in the derivations in this Appendix. The IBP relations and the differential equations have been automated by, e.g. the packages~\cite{Lee:2012cn, Smirnov:2019qkx}, which can be used to calculate arbitrary $n$-point function of theory \ibp defined in Section~\ref{sec:Unitary}. We have verified that our analytic calculations match the results produced by \texttt{LiteRed2}~\cite{Lee:2012cn}.

\subsection*{The IBP relations of $n$-point functions}

The integral we wish to evaluate always takes the form:
\begin{equation}\label{eq:loopint}
    J^{(s)}(n_0,n_1,\cdots,n_s) := \int d^dl j^{(s)}(n_0,n_1,\cdots,n_s) = \int d^dl \frac{1}{(l^2+1)^{n_0}} \prod_{i=1}^s \frac{1}{(l\cdot p_i)^{n_i}}
\end{equation}
with one ``loop momentum'' $l$ and $E$ external legs $p_i$. We note that
\begin{equation}
    J^{(s)}(1,\cdots,1) \equiv G^{(s)}(X_1,\cdots,X_s)
\end{equation}
after identifying $l$ with $\Theta$ and each $p_i$ with $X_i$. The form of~(\ref{eq:loopint}) suggests us to make use of the IBP method to find relations among different $J^{(s)}(n_0,n_1,\cdots,n_s)$'s in order to systematically evaluate the integrals.

We define the differential operator:
\begin{equation}
    \mathcal{D}_{q} F := \int d^dl \frac{\partial}{\partial l} \cdot (q f)
\end{equation}
for $q = l,p_1,\cdots,p_s$ and $F = \int d^dl f$. The IBP relations are derived from the fact:
\begin{equation}
    \mathcal{D}_q J^{(s)}(n_0,n_1,\cdots,n_s) = 0,\ \forall q.
\end{equation}
In the derivation we impose the null-cone condition $p_i^2 = 0,\ \forall i$.

\paragraph{IBP of 0-point function}

We consider the IBP of $J^{(0)}(n_0)$. We have:
\begin{equation}
    \mathcal{D}_l J^{(0)} 
    = 2n_0 J^{(0)}(n_0 + 1) + (d-2n_0) J^{(0)}(n_0).
\end{equation}
Therefore the IBP relation is:
\begin{equation}
    J^{(0)}(n_0 + 1) = \frac{2n_0 - d}{2n_0} J^{(0)}(n_0).
\end{equation}
The master integral is $J^{(0)}(n_0)$.

\paragraph{IBP of 1-point function}

We consider the IBP for $J^{(1)}(n_0,n_1)$. We have:
\begin{equation}
    \begin{split}
        \mathcal{D}_l J^{(1)} 
        &= 2n_0 J^{(1)}(n_0+1,n_1) + (d-2n_0-n_1) J^{(1)}(n_0,n_1), \\
        \mathcal{D}_{p_1} J^{(1)} p_1^2) \\
        &= -2n_0 J^{(1)}(n_0+1, n_1-1).
    \end{split}
\end{equation}
Therefore the IBP relations are:
\begin{equation}
    \begin{split}
        J^{(1)}(n_0+1,n_1) &= \frac{2n_0+n_1-d}{2n_0} J^{(1)}(n_0,n_1), \\
        2n_0 J^{(1)}(n_0+1, n_1-1) &= 0.
    \end{split}
\end{equation}
The master integral is $J^{(1)}(1,n_1)$.

\paragraph{IBP of 2-point function}

We consider the IBP of $J^{(2)}(n_0,n_1,n_2)$. We have:
\begin{equation}
    \begin{split}
        \mathcal{D}_l J^{(2)} &= 2n_0 J^{(2)}(n_0+1,n_1,n_2) + (d-2n_0-n_1-n_2) J^{(2)}(n_0,n_1,n_2), \\
        \mathcal{D}_{p_1} J^{(2)} &= -2n_0 J^{(2)}(n_0+1, n_1-1, n_2) - n_2 (p_1\cdot p_2) J^{(2)}(n_0, n_1, n_2+1), \\
        \mathcal{D}_{p_2} J^{(2)} &= -2n_0 J^{(2)}(n_0+1, n_1, n_2-1) - n_1 (p_1\cdot p_2) J^{(2)}(n_0, n_1+1, n_2).
    \end{split}
\end{equation}
Therefore the IBP relations are:
\begin{equation}
    2n_0 J^{(2)}(n_0+1,n_1,n_2) = (2n_0 + n_1 + n_2 - d) J^{(2)}(n_0,n_1,n_2) = -n_2(p_1\cdot p_2) J^{(2)}(n_0, n_1+1, n_2+1).
\end{equation}
The following relation is particularly interesting:
\begin{equation}
    J^{(2)}(n_0, n_1+1, n_2+1) = \frac{d - 2n_0 - n_1 - n_2}{n_2(p_1\cdot p_2)} J^{(2)}(n_0, n_1, n_2) = \frac{d - 2n_0 - n_1 - n_2}{n_1(p_1\cdot p_2)} J^{(2)}(n_0, n_1, n_2),
\end{equation}
which further leads to:
\begin{equation}
    \frac{1}{n_2} J^{(2)}(n_0,n_1,n_2) = \frac{1}{n_1} J^{(2)}(n_0,n_1,n_2)
\end{equation}
which, when $n_1\neq n_2$, is consistent only if $J(n_0, n_1, n_2) = 0$. This implies:
\begin{equation}\label{eq:vanishing_diffweight_2pt}
    J^{(2)}(1,n_1,n_2) = \int d^dl \frac{1}{l^2 + 1} \frac{1}{(l\cdot p_1)^{n_1}} \frac{1}{(l\cdot p_2)^{n_2}} = 0,\ n_1\neq n_2
\end{equation}
which in our original setup means that the 2-point function of two fields with different weights vanishes.

\paragraph{IBP of 3-point function}

We consider the IBP of $J^{(3)}(n_0, n_1, n_2, n_3)$. We have:
\begin{equation}
    \begin{split}
        \mathcal{D}_l J^{(3)} &= 2n_0 J^{(3)}(n_0+1,n_1,n_2,n_3) + (d-2n_0-n_1-n_2-n_3) J^{(3)}(n_0,n_1,n_2,n_3), \\
        \mathcal{D}_{p_1} J^{(3)} &= -2n_0 J^{(3)}(n_0+1,n_1-1,n_2,n_3) - n_2(p_1\cdot p_2) J^{(3)}(n_0,n_1,n_2+1,n_3) \\
        &\quad - n_3(p_1\cdot p_3) J^{(3)}(n_0,n_1,n_2,n_3+1), \\
        \mathcal{D}_{p_2} J^{(3)} &=-2n_0 J^{(3)}(n_0+1,n_1,n_2-1,n_3) - n_1(p_1\cdot p_2) J^{(3)}(n_0,n_1+1,n_2,n_3) \\
        &\quad - n_3(p_2\cdot p_3) J^{(3)}(n_0,n_1,n_2,n_3+1), \\
        \mathcal{D}_{p_3} J^{(3)} &=-2n_0 J^{(3)}(n_0+1,n_1,n_2,n_3-1) - n_1(p_1\cdot p_3) J^{(3)}(n_0,n_1+1,n_2,n_3) \\
        &\quad - n_2(p_2\cdot p_3) J^{(3)}(n_0,n_1,n_2+1,n_3).
    \end{split}
\end{equation}
The relation from $\mathcal{D}_l J^{(3)} = 0$ is:
\begin{equation}
    J^{(3)}(n_0+1,n_1,n_2,n_3) = \frac{2n_0 + n_1 + n_2 + n_3 - d}{2n_0} J^{(3)}(n_0,n_1,n_2,n_3)
\end{equation}
which leads to master integral $J^{(3)}(1,n_1,n_2,n_3)$. The relation from $\mathcal{D}_{p_1} J^{(3)} = \mathcal{D}_{p_2} J^{(3)} = \mathcal{D}_{p_3} J^{(3)} = 0$ gives rise to:
\begin{equation}\label{eq:IBPJ3}
    \begin{split}
        & n_2(p_1\cdot p_2) J^{(3)}(n_0,n_1+1,n_2+1,n_3) + n_3(p_1\cdot p_3) J^{(3)}(n_0,n_1+1,n_2,n_3+1) \\
        =\ & n_1(p_1\cdot p_2) J^{(3)}(n_0,n_1+1,n_2+1,n_3) + n_3(p_2\cdot p_3) J^{(3)}(n_0,n_1,n_2+1,n_3+1) \\
        =\ & n_1(p_1\cdot p_3) J^{(3)}(n_0,n_1+1,n_2,n_3+1) + n_2(p_2\cdot p_3) J^{(3)}(n_0,n_1,n_2+1,n_3+1).
    \end{split}
\end{equation}

\bigskip

The IBP relations for $J^{(s)}$ with $s > 3$ can be obtained in a similar fashion, which we will introduce the reader to the package~\cite{Lee:2012cn, Smirnov:2019qkx} for automation.

\subsection*{Differential equations of $n$-point functions}

Having obtained the IBP relations which leads us to the master integrals, we now derive the differential equations that are satisfied by the master integrals.

From the chain rule we have:
\begin{equation}
    \frac{\partial f}{\partial p_{im}} = \frac{\partial f}{\partial (p_i \cdot p_j)} \frac{\partial (p_i \cdot p_j)}{\partial p_{im}} = \sum_j p_{jm} \frac{\partial f}{\partial (p_i\cdot p_j)}
\end{equation}
where $p_{im}$ is the $m^{\text{th}}$ component of $p_i$. Here we have made the key assumption that $f$ is a function of all $p_i\cdot p_j$'s. The above chain rule relation can be written into the more convenient matrix form as follows:
\begin{equation}
    \begin{pmatrix}
        \frac{\partial}{\partial p_{11}} & \cdots & \frac{\partial}{\partial p_{1d}} \\
        \vdots & \ddots & \vdots \\
        \frac{\partial}{\partial p_{s1}} & \cdots & \frac{\partial}{\partial p_{sd}}
    \end{pmatrix} f = \begin{pmatrix}
        \frac{\partial}{\partial (p_1\cdot p_1)} & \cdots & \frac{\partial}{\partial (p_1\cdot p_s)} \\
        \vdots & \ddots & \vdots \\
        \frac{\partial}{\partial (p_s\cdot p_1)} & \cdots & \frac{\partial}{\partial (p_s\cdot p_s)}
    \end{pmatrix} f \cdot \begin{pmatrix}
            p_{11} & \cdots & p_{1d} \\
            \vdots & \ddots & \vdots \\
            p_{s1} & \cdots & p_{sd}
    \end{pmatrix}.
\end{equation}
Multiplying both sides by $(p_{im})^T$, we get:
\begin{equation}
    \begin{split}
        \begin{pmatrix}
        \frac{\partial}{\partial p_{11}} & \cdots & \frac{\partial}{\partial p_{1d}} \\
        \vdots & \ddots & \vdots \\
        \frac{\partial}{\partial p_{s1}} & \cdots & \frac{\partial}{\partial p_{sd}}
    \end{pmatrix} f \begin{pmatrix}
            p_{11} & \cdots & p_{s1} \\
            \vdots & \ddots & \vdots \\
            p_{1d} & \cdots & p_{sd}
    \end{pmatrix} 
    &= \begin{pmatrix}
        \frac{\partial}{\partial (p_1\cdot p_1)} & \cdots & \frac{\partial}{\partial (p_1\cdot p_s)} \\
        \vdots & \ddots & \vdots \\
        \frac{\partial}{\partial (p_s\cdot p_1)} & \cdots & \frac{\partial}{\partial (p_s\cdot p_s)}
    \end{pmatrix} f \cdot \mathbb{G}
    \end{split}
\end{equation}
where $\mathbb{G}_{ij} = p_i\cdot p_j$. Since $\mathbb{G}$ is generically invertible, we multiply both sides of the above equation by $\mathbb{G}^{-1}$ to get:
\begin{equation}
    \begin{split}
        \begin{pmatrix}
        \frac{\partial}{\partial (p_1\cdot p_1)} & \cdots & \frac{\partial}{\partial (p_1\cdot p_s)} \\
        \vdots & \ddots & \vdots \\
        \frac{\partial}{\partial (p_s\cdot p_1)} & \cdots & \frac{\partial}{\partial (p_s\cdot p_s)}
    \end{pmatrix}
    &= \begin{pmatrix}
        \frac{\partial}{\partial p_{11}} & \cdots & \frac{\partial}{\partial p_{1d}} \\
        \vdots & \ddots & \vdots \\
        \frac{\partial}{\partial p_{s1}} & \cdots & \frac{\partial}{\partial p_{sd}}
    \end{pmatrix} f \begin{pmatrix}
        \sum_i p_{i1}\mathbb{G}^{-1}_{i1} & \cdots & \sum_i p_{i1}\mathbb{G}^{-1}_{is} \\
        \vdots & \ddots & \vdots \\
        \sum_i p_{id}\mathbb{G}^{-1}_{i1} & \cdots & \sum_i p_{id}\mathbb{G}^{-1}_{is}
    \end{pmatrix}.
    \end{split}
\end{equation}
Writing in terms of components we arrive at:
\begin{equation}\label{eq:PD}
    \frac{\partial f}{\partial (p_i\cdot p_j)} = \sum_n \frac{\partial f}{\partial p_{in}} \sum_k p_{kn} \mathbb{G}^{-1}_{kj} = \sum_k \mathbb{G}^{-1}_{kj} p_k \cdot \frac{\partial f}{\partial p_i}
\end{equation}
where we have turned $\sum_n$ into dot product between $d$-dimensional vectors. The set of differential equations~(\ref{eq:PD}) is satisfied by any $n$-point function $f$. In our setup

\paragraph{2-point function}

We consider the 2-point function:
\begin{equation}
    J^{(2)}(1,n_1,n_2) = \int d^dl \frac{1}{l^2 + 1} \frac{1}{(l\cdot p_1)^{n_1}} \frac{1}{(l\cdot p_2)^{n_2}}.
\end{equation}
Due to~(\ref{eq:vanishing_diffweight_2pt}) we focus on the $n_1 = n_2$ cases. Let us first consider $\frac{\partial J^{(2)}}{\partial (p_1\cdot p_1)}$. We have:
\begin{equation}
    \begin{split}
        \frac{\partial J^{(2)}(1,n,n)}{\partial (p_1\cdot p_1)} &= -\frac{n}{p_1\cdot p_2} J^{(2)}(1,n+1,n-1) = 0
    \end{split}
\end{equation}
due to~(\ref{eq:vanishing_diffweight_2pt}). Similarly one can show that $\frac{\partial J^{(2)}(1,n,n)}{\partial (p_2\cdot p_2)} = 0$. Therefore $J^{(2)}(1,n,n)$ does not depend on any $p_i\cdot p_i$.

We then study $\frac{\partial J^{(2)}(1,n,n)}{\partial (p_1\cdot p_2)}$. We have:
\begin{equation}
    \frac{\partial J^{(2)}(1,n,n)}{\partial (p_1\cdot p_2)} = - \frac{n}{p_1\cdot p_2} J^{(2)}(1,n,n).
\end{equation}
Therefore for the 2-point function we have:
\begin{equation}\label{eq:J_2pt}
    J^{(2)}(1,n,n) = \frac{C}{(p_1\cdot p_2)^n}
\end{equation}
with integral constant $C$.

\paragraph{3-point function}

We consider the 3-point function:
\begin{equation}
    J^{(2)}(1,n_1,n_2,n_3) = \int d^dl \frac{1}{l^2 + 1} \frac{1}{(l\cdot p_1)^{n_1}} \frac{1}{(l\cdot p_2)^{n_2}} \frac{1}{(l\cdot p_3)^{n_3}}.
\end{equation}
Let us first consider $\frac{\partial J^{(3)}}{\partial (p_1\cdot p_1)}$. We have:
\begin{equation}\label{eq:J0123}
    \begin{split}
        \frac{\partial J^{(3)}(1,n_1,n_2,n_3)}{\partial (p_1\cdot p_1)} &= \frac{ n_1 (p_2\cdot p_3)^2 }{\mathcal{D}} J^{(3)}(1,n_1,n_2,n_3) - \frac{ n_1(p_1\cdot p_3)(p_2\cdot p_3) }{\mathcal{D}} J^{(3)}(1,n_1+1,n_2-1,n_3) \\
        &\quad - \frac{ n_1(p_1\cdot p_2)(p_2\cdot p_3) }{\mathcal{D}} J^{(3)}(1,n_1+1,n_2,n_3-1) 
    \end{split}
\end{equation}
where $\mathcal{D} = 2(p_1\cdot p_2)(p_1\cdot p_3)(p_2\cdot p_3)$. Here we focus only on the case $n_1 = n_2 = n_3 = 1$. Due to:
\begin{equation}
    J^{(3)}(1,2,0,1) = J^{(3)}(1,2,1,0) = J^{(2)}(1,2,1) = 0
\end{equation}
with suitable relabeling of external legs, (\ref{eq:J0123}) simplifies to:
\begin{equation}
    \frac{\partial J^{(3)}(1,1,1,1)}{\partial (p_1\cdot p_1)} = \frac{ (p_2\cdot p_3) }{2(p_1\cdot p_2)(p_1\cdot p_3)} J^{(3)}(1,1,1,1)
\end{equation}
the solution of which is:
\begin{equation}
    J^{(3)}(1,1,1,1) = C e^{p_1\cdot p_1}
\end{equation}
with $C = C(p_i\cdot p_j)$, $(i,j) \neq (1,1)$. Imposing the null-cone constraint $p_i^2 = 0$ we have $J^{(3)}(1,1,1,1)$, which implies that $J^{(3)}(1,1,1,1)$ does not depend on $p_1^2$. Similarly one can show that $J^{(3)}(1,1,1,1)$ does not depend on any $p_i^2$.

We then study $\frac{\partial J^{(3)}(1,1,1,1)}{\partial (p_1\cdot p_2)}$. After making use of the fact $J^{(3)}(1,2,0,1) = J^{(3)}(1,2,1,0) = 0$, we have:
\begin{equation}
    \frac{\partial J^{(3)}(1,1,1,1)}{\partial (p_1\cdot p_2)} = -\frac{(p_1\cdot p_3)(p_2\cdot p_3)}{\mathcal{D}} J^{(3)}(1,1,1,1) = -\frac{1}{2(p_1\cdot p_2)} J^{(3)}(1,1,1,1).
\end{equation}
The solution of the above equation is:
\begin{equation}
    J^{(3)}(1,1,1,1) = \frac{C}{\sqrt{p_1\cdot p_2}}
\end{equation}
with $C = C(p_1\cdot p_3, p_2\cdot p_3)$. Similarly, after making use of the differentials $\frac{\partial J^{(3)}(1,1,1,1)}{\partial (p_1\cdot p_3)}$ and $\frac{\partial J^{(3)}(1,1,1,1)}{\partial (p_2\cdot p_3)}$, we arrive at:
\begin{equation}
\label{eqn:app_3pt}
    J^{(3)}(1,1,1,1) = \frac{C}{\sqrt{(p_1\cdot p_2)(p_1\cdot p_3)(p_2\cdot p_3)}}
\end{equation}
with constant $C$. By symmetry in this $(1,1,1,1)$ case, we must have $C=0$, but it is satisfying to see that the Lorentz structure of \eqref{eqn:app_3pt} gives the correct conformal structure required of three point functions upon restriction to the Poincar\'e section.

\paragraph{4-point function}

We consider the 4-point function:
\begin{equation}
    J^{(4)}(1,n_1,n_2,n_3,n_4) = \int d^dl \frac{1}{l^2 + 1} \frac{1}{(l\cdot p_1)^{n_1}} \frac{1}{(l\cdot p_2)^{n_2}} \frac{1}{(l\cdot p_3)^{n_3}} \frac{1}{(l\cdot p_4)^{n_4}}.
\end{equation}
For simplicity we write $p_{ij} := p_i\cdot p_j$ and define:
\begin{equation}
    \mathcal{D} = p_{12}^2p_{34}^2 + p_{13}^2p_{24}^2 + p_{14}^2p_{23}^2 - 2p_{12}p_{34} p_{13}p_{24} -  2p_{12}p_{34} p_{14}p_{23} - 2 p_{13}p_{24} p_{14}p_{23}.
\end{equation}
We again focus on $J^{(4)}(1,1,1,1,1)$ for which we have:
\begin{equation}\label{eq:J11111a}
    \begin{split}
        \frac{\partial J^{(4)}(1,1,1,1,1)}{\partial p_{11}} &= - \frac{2(p_{23}p_{24}p_{34})}{\mathcal{D}} J^{(4)}(1,1,1,1,1) + \frac{p_{34}(p_{14}p_{23} + p_{13}p_{24} - p_{12}p_{34})}{\mathcal{D}} J^{(4)}(1,2,0,1,1) \\
        &\quad + \frac{p_{24}(p_{14}p_{23} + p_{12}p_{34} - p_{13}p_{24})}{\mathcal{D}} J^{(4)}(1,2,1,0,1) \\
        &\quad + \frac{p_{23}(p_{12}p_{34} + p_{13}p_{24} - p_{14}p_{23})}{\mathcal{D}} J^{(4)}(1,2,1,1,0).
    \end{split}
\end{equation}
We now have to study terms like $J^{(4)}(1,2,1,1,0) = J^{(3)}(1,2,1,1)$ appearing in the above equation. Recall from~(\ref{eq:J0123}) we have:
\begin{equation}
    \begin{split}
        \frac{\partial J^{(3)}(1,2,1,1)}{\partial (p_{11})} &= \frac{ 2 (p_2\cdot p_3)^2 }{\mathcal{D}} J^{(3)}(1,2,1,1) - \frac{ 2(p_1\cdot p_3)(p_2\cdot p_3) }{\mathcal{D}} J^{(3)}(1,3,0,1) \\
        &\quad - \frac{ 2(p_1\cdot p_2)(p_2\cdot p_3) }{\mathcal{D}} J^{(3)}(1,3,1,0) \\
        &= -\frac{2p_{23}}{p_{12}p_{13}} J^{(3)}(1,2,1,1)
    \end{split}
\end{equation}
which implies that $J^{(3)}(1,2,1,1)$, hence $J^{(4)}(1,2,1,1,0)$, does not depend on $p_{11}$ at all after imposing the $p_{11} = 0$ condition. This in turn implies that $J^{(4)}(1,1,1,1,1)$ does not depend on $p_{11}$ via~(\ref{eq:J11111a}) after imposing $p_{11} = 0$. Similarly one can show that $J^{(4)}(1,1,1,1,1)$ does not depend on any $p_{ii}$.

We now turn to $\frac{\partial J^{(4)}(1,1,1,1,1)}{\partial p_{12}}$. We have:
\begin{equation}\label{eq:J11111b}
    \begin{split}
        \frac{\partial J^{(4)}(1,1,1,1,1)}{\partial p_{12}} &= \frac{p_{34}(p_{14}p_{23} + p_{13}p_{24} - p_{12}p_{34})}{\mathcal{D}} J^{(4)}(1,1,1,1,1) - \frac{2p_{13}p_{14}p_{34}}{\mathcal{D}} J^{(4)}(1,2,0,1,1) \\
        &\quad + \frac{p_{14}(p_{12}p_{34} + p_{13}p_{24} - p_{14}p_{23})}{\mathcal{D}} J^{(4)}(1,2,1,0,1) \\
        &\quad + \frac{p_{13}(p_{12}p_{34} + p_{14}p_{23} - p_{13}p_{24})}{\mathcal{D}} J^{(4)}(1,2,1,1,0).
    \end{split}
\end{equation}
Clearly we have to calculate $J^{(3)}(1,2,1,1)$ for which we have:
\begin{equation}
    \begin{split}
        \frac{\partial J^{(3)}(1,2,1,1)}{\partial p_{12}} &= -\frac{1}{p_{12}} J^{(3)}(1,2,1,1), \\
        \frac{\partial J^{(3)}(1,2,1,1)}{\partial p_{13}} &= -\frac{1}{p_{13}} J^{(3)}(1,2,1,1), \\
        \frac{\partial J^{(3)}(1,2,1,1)}{\partial p_{23}} &= -\frac{1}{p_{13}} J^{(3)}(1,1,2,1) - \frac{1}{p_{23}} J^{(3)}(1,2,1,1) + \frac{p_{12}}{p_{13}p_{23}} J^{(3)}(1,2,2,0)
    \end{split}
\end{equation}
The first two equations lead to:
\begin{equation}\label{eq:J3_p12p13}
    J^{(3)}(1,2,1,1) = \frac{C(p_{23})}{p_{12}p_{13}}
\end{equation}
and using~(\ref{eq:J_2pt}) the last equation becomes:
\begin{equation}\label{eq:last3pt}
     \frac{\partial J^{(3)}(1,2,1,1)}{\partial p_{23}} = -\frac{1}{p_{13}} J^{(3)}(1,1,2,1) - \frac{1}{p_{23}} J^{(3)}(1,2,1,1) + \frac{C}{p_{12}p_{13}p_{23}}
\end{equation}
with constant $C$. With $(n_0,n_1,n_2,n_3) = (1,1,1,0)$ the IBP relation~(\ref{eq:IBPJ3}) becomes:
\begin{equation}\label{eq:IBP_J1211_more}
    p_{13} J^{(3)}(1,2,1,1) + p_{23} J^{(3)}(1,1,2,1) = p_{12} J^{(3)}(1,2,1,0) = p_{12} J^{(2)}(1,2,1) = 0.
\end{equation}
Plugging the above result into~(\ref{eq:last3pt}), we have:
\begin{equation}
    \frac{\partial J^{(3)}(1,2,1,1)}{\partial p_{23}} = \frac{C}{p_{12}p_{13}p_{23}}.
\end{equation}
Combining with~(\ref{eq:J3_p12p13}) we have:
\begin{equation}
    J^{(3)}(1,2,1,1) = \frac{C \log(p_{23})}{p_{12}p_{13}}.
\end{equation}
Similarly we have:
\begin{equation}
    J^{(3)}(1,1,2,1) = \frac{C \log(p_{13})}{p_{12}p_{23}}.
\end{equation}
The IBP relation~(\ref{eq:IBP_J1211_more}) fixes $C = 0$. Therefore we have:
\begin{equation}
    J^{(3)}(1,2,1,1) = 0.
\end{equation}

Given $J^{(3)}(1,2,1,1) = 0$,~(\ref{eq:J11111b}) becomes:
\begin{equation}
    \frac{\partial J^{(4)}(1,1,1,1,1)}{\partial p_{12}} = \frac{p_{34}(p_{14}p_{23} + p_{13}p_{24} - p_{12}p_{34})}{\mathcal{D}} J^{(4)}(1,1,1,1,1).
\end{equation}
It is not hard to check that the coefficient in front of $J^{(4)}$ on the RHS of the above equation is equal to:
\begin{equation}
    \frac{\partial}{\partial p_{12}} \left( C_1' - \frac{1}{2}\log (\mathcal{D}) \right)
\end{equation}
with $C_1' = C_1'(p_{ij})$ which depends on all $p_{ij}$ for $i < j$ except $p_{12}$. Therefore, the solution of~(\ref{eq:J11111b}) is:
\begin{equation}\label{eq:J11111_p12}
    J^{(4)}(1,1,1,1,1) = \frac{C_1}{\sqrt{\mathcal{D}}} 
\end{equation}
where $C_1 = \exp(C_1')$.

Similarly we consider the differential equation:
\begin{equation}
    \frac{\partial J^{(4)}(1,1,1,1,1)}{\partial p_{23}} = \frac{\partial}{\partial p_{23}} \left( C_2' - \frac{1}{2}\log (\mathcal{D}) \right) J^{(4)}(1,1,1,1,1)
\end{equation}
with $C_2' = C_2'(p_{ij})$ which depends on all $p_{ij}$ for $i < j$ except $p_{23}$. This leads to the solution:
\begin{equation}\label{eq:J11111_p23}
    J^{(4)}(1,1,1,1,1) = \frac{C_2}{\sqrt{\mathcal{D}}} 
\end{equation}
where $C_2 = \exp(C_2')$. Consistency of solutions~(\ref{eq:J11111_p12}) and~(\ref{eq:J11111_p23}) dictates $C_1 = C_2 = C$ and $C$ does not depend on either $p_{12}$ or $p_{23}$.

The other partial derivatives with respect to $p_{ij}$'s can be performed in exactly the same manner and we conclude that the integral constant $C$ is indeed a number, i.e. we have:
\begin{equation}
    J^{(4)}(1,1,1,1,1) = \frac{C}{\sqrt{\mathcal{D}}}
\end{equation}
with constant $C$.

\section{The Energy Momentum Tensor}
\label{app:stress-tensor}

The free boson is taken to be a large $N\rightarrow\infty$ ensemble of fields defined as
\begin{equation}\label{eq:small-Phi}
\varphi(X) = \frac{1}{\sqrt{N}} \sum_{i=1}^{N} w_i \Phi_i(X), \quad \quad X\in \bR^{D+1,1},
\end{equation}
where 
\begin{equation}\label{eq:cap-Phi}
    \Phi_i(X)=\frac{1}{\theta_{i}\cdot X}
\end{equation}
The four point function $\langle \varphi(X_1) \varphi(X_2) \varphi(X_3) \varphi(X_4)\rangle$ in the $N\rightarrow\infty$ limit can be written as,
\begin{equation}
    \langle \varphi(X_1) \varphi(X_2) \varphi(X_3) \varphi(X_4)\rangle=\frac{g(u,v)}{x_{12}^2 x_{34}^2},
\end{equation}
where using equation \ref{eq:g_mixing} for $\Delta=1$, $g(u,v)$ is given by
\begin{equation}
    g(u,v)=1+u+\frac{u}{v},
    \end{equation}
whose conformal block decomposition gives us twist $2$ operators \cite{Rattazzi:2008pe}, including the energy-momentum tensor. The energy-momentum tensor in the 4d free field theory is given by
\begin{equation}\label{eq:Tmunu}
 T_{\mu\nu}(X) = \varphi(X) \partial_\mu \partial_\nu \varphi(X) - 2\left[\partial_\mu \varphi(X) \partial_\nu \varphi(X) - \frac14 \delta_{\mu\nu} (\partial \varphi(X))^2 \right]
\end{equation}
We will calculate the two-point correlator.
\begin{equation}\label{eq:2ptT}
\begin{aligned}
        \langle T_{\mu\nu}(X) T_{\rho\sigma}(Y) \rangle =& \langle \left(\varphi(X) \partial_\mu \partial_\nu \varphi(X) - 2\left[\partial_\mu \varphi(X) \partial_\nu \varphi(X) - \frac14 \delta_{\mu\nu} (\partial \varphi(X))^2 \right]\right) \\
        & \left(\varphi(Y) \partial_\sigma \partial_\rho \varphi(Y) - 2\left[\partial_\rho \varphi(Y) \partial_\sigma \varphi(Y) - \frac14 \delta_{\rho\sigma} (\partial \varphi(Y))^2 \right] \right) \rangle
\end{aligned}
\end{equation}
Expanding all the terms in \ref{eq:2ptT}, we get
\begin{equation}\label{eq:2ptTfinal}
\begin{aligned}
        &\langle T_{\mu\nu}(X) T_{\rho\sigma}(Y) \rangle \\
        &= \langle \varphi(X) \partial_\mu \partial_\nu \varphi(X) \varphi(Y) \partial_\sigma \partial_\rho \varphi(Y) \rangle -2 \langle \varphi(X) \partial_\mu \partial_\nu \varphi(X) \partial_\rho \varphi(Y) \partial_\sigma \varphi(Y)\rangle + \frac{1}{2} \langle \varphi(X) \partial_\mu \partial_\nu \varphi(X) \delta_{\rho\sigma} (\partial \varphi(Y))^2 \rangle \\
        & -2 \langle \partial_\mu \varphi(X) \partial_\nu \varphi(X) \varphi(Y) \partial_\sigma \partial_\rho \varphi(Y) \rangle+ 4 \langle \partial_\mu \varphi(X) \partial_\nu \varphi(X) \partial_\rho \varphi(Y) \partial_\sigma \varphi(Y)\rangle - \langle \partial_\mu \varphi(X) \partial_\nu \varphi(X) \delta_{\rho\sigma} (\partial \varphi(Y))^2 \rangle \\
        & + \frac{1}{2} \delta_{\mu\nu} \langle (\partial \varphi(X))^2 \varphi(Y) \partial_\sigma \partial_\rho \varphi(Y) \rangle - \delta_{\mu\nu} \langle (\partial \varphi(X))^2 \partial_\rho \varphi(Y) \partial_\sigma \varphi(Y)\rangle + \frac{1}{4} \delta_{\mu\nu} \langle (\partial \varphi(X))^2  \delta_{\rho\sigma} (\partial \varphi(Y))^2 \rangle
\end{aligned}
\end{equation}
Rewriting \ref{eq:2ptT} as an expectation value in the parameter space, using the definitions in \ref{eq:small-Phi}, \ref{eq:cap-Phi}, we get
\begin{equation}\label{eq:2ptTexp}
\begin{aligned}
       \langle T_{\mu\nu}(X) T_{\rho\sigma}(Y) \rangle = \langle \sum_{i,j,k,l=1}^{N} & w_i w_j w_k w_l   \left(\Phi_i(X) \partial_\mu \partial_\nu \Phi_j(X) - 2\left[\partial_\mu \Phi_i(X) \partial_\nu \Phi_j(X) - \frac14 \delta_{\mu\nu} \partial \Phi_i(X)\partial \Phi_j(X) \right]\right) \\
        & \left(\Phi_k(Y) \partial_\sigma \partial_\rho\Phi_l(Y) - 2\left[\partial_\rho \Phi_k(Y) \partial_\sigma \Phi_l(Y) - \frac14 \delta_{\rho\sigma} \partial \Phi_k(X)\partial \Phi_l(X) \right] \right)\rangle \frac{1}{N^2}
\end{aligned}
\end{equation}
However, based on our calculations in section \ref{sec:freeboson}, we know from the expectation value of $\langle w_i w_j w_k w_l \rangle := \int d\omega P(\omega) w_i w_j w_k w_l$, that
\begin{equation}\label{eq:4pt-omega}
\begin{aligned}
   \frac{1}{N^2} \sum_{i,j,k,l=1}^N \langle w_i w_j w_k w_l \rangle &= \frac{1}{N^2}\left( \sum_{i,j,k,l=1}^N \gamma^4 \delta_{ij} \delta_{jk} \delta_{kl} + \sigma^4 \left( \sum_{i,j} \sum_{k,l\neq i} \delta_{ij}\delta_{kl} + \sum_{i,k} \sum_{j,l\neq i} \delta_{ik}\delta_{jl} + \sum_{i,l} \sum_{jk\neq i} \delta_{il}\delta_{jk} \right) \right)\\
   &= \frac{\gamma^4}{N} + \sigma^4 \left(1-\frac{1}{N}\right) \left(\delta_{ij}\delta_{kl}+\delta_{ik}\delta_{jl}+\delta_{il}\delta_{jk}\right)\\
   & \longrightarrow_{\text{lim}_{N\rightarrow\infty}} \sigma^4 \left(\delta_{ij}\delta_{kl}+\delta_{ik}\delta_{jl}+\delta_{il}\delta_{jk}\right)
\end{aligned}
\end{equation}
In the limit $N\rightarrow\infty$, we only have to worry about the terms involving $\sigma^4$ in \ref{eq:4pt-omega} i.e., we only keep the relevant terms that do not vanish at $N\rightarrow\infty$. In the equation \ref{eq:2ptTfinal} there are nine terms in total. We will evaluate them explicitly. Note that from now on the expectation $\langle . \rangle$ refers to the expectation values with respect to all the $\theta_i$s (as the $\omega$ integral has been already performed which gave us the index structure in $i,j,k,l$). For example, let us consider the term $\langle \partial_\mu \varphi(X) \partial_\nu \varphi(X) \partial_\rho \varphi(Y) \partial_\sigma \varphi(Y)\rangle$,
\begin{equation}\label{eq:example-term-calc}
    \begin{aligned}
        & \langle \partial_{\mu}\varphi(X)\partial_{\nu}\varphi(X) \partial_{\rho}\varphi(Y)\partial_{\sigma}\varphi(Y)\rangle \\
        & = \sum_{i,j,k,l}\frac{\sigma^4}{N^2} \langle \frac{\theta_{i\mu}\theta_{j\nu}\theta_{k\rho}\theta_{l\sigma}}{(\theta_i \cdot X)^2 (\theta_j \cdot X)^2 (\theta_k \cdot Y)^2 (\theta_l \cdot Y)^2} \left(\delta_{ij}\delta_{kl}+\delta_{ik}\delta_{jl}+\delta_{il}\delta_{jk}\right) \rangle \\
        &= \sigma^4 \int d\theta_i d\theta_k P(\theta_i)P(\theta_k) \frac{\theta_{i\mu}\theta_{i\nu}\theta_{k\rho}\theta_{k\sigma}}{(\theta_i \cdot X)^4 (\theta_k \cdot Y)^4}\\
        & + \sigma^4 \int d\theta_i d\theta_k P(\theta_i)P(\theta_k)\frac{\theta_{i\mu}\theta_{k\nu}\theta_{i\rho}\theta_{k\sigma}}{(\theta_i \cdot X)^2 (\theta_i \cdot Y)^2 (\theta_k \cdot X)^2 (\theta_k \cdot Y)^2}\\
        & + \sigma^4 \int d\theta_i d\theta_k\frac{\theta_{i\mu}\theta_{k\nu}\theta_{k\rho}\theta_{i\sigma}}{(\theta_i \cdot X)^2 (\theta_i \cdot Y)^2 (\theta_k \cdot X)^2 (\theta_k \cdot Y)^2}\\
        & = \frac{\sigma^4}{36} \left( \partial_{\mu} \partial_{\nu} \int d\theta_i P(\theta_i) \frac{1}{(\theta_i \cdot X)^2}\right) \left( \partial_{\rho} \partial_{\sigma} \int d\theta_k P(\theta_k) \frac{1}{(\theta_k \cdot Y)^2}\right)\\
        & + \sigma^4 \left( \partial_{\mu} \partial_{\rho} \int d\theta_i P(\theta_i) \frac{1}{(\theta_i \cdot X)(\theta_i \cdot Y)}\right) \left( \partial_{\nu} \partial_{\sigma} \int d\theta_k P(\theta_k) \frac{1}{(\theta_k \cdot X)(\theta_k \cdot Y)}\right)\\
        & + \sigma^4 \left( \partial_{\mu} \partial_{\sigma} \int d\theta_i P(\theta_i) \frac{1}{(\theta_i \cdot X)(\theta_i \cdot Y)}\right) \left( \partial_{\nu} \partial_{\rho} \int d\theta_k P(\theta_k) \frac{1}{(\theta_k \cdot X)(\theta_k \cdot Y)}\right)\\
        &= \frac{\sigma^4}{36} \left( \partial_{\mu} \partial_{\nu} G^{1}_{\Delta=2}(X) \right) \left( \partial_{\rho} \partial_{\sigma} G^{1}_{\Delta=2}(Y) \right)\\
        & + \sigma^4\left[\left( \partial_{\mu} \partial_{\rho}  G^{2}_{\Delta=1}(X,Y) \right) \left(\partial_{\nu}\partial_{\sigma}  G^{2}_{\Delta=1}(X,Y) \right) + \left( \partial_{\mu} \partial_{\sigma}  G^{2}_{\Delta=1}(X,Y) \right) \left(\partial_{\nu}\partial_{\rho}  G^{2}_{\Delta=1}(X,Y) \right) \right]\\
        & = \sigma^4\left[\left( \partial_{\mu} \partial_{\rho}  G^{2}_{\Delta=1}(X,Y) \right) \left(\partial_{\nu}\partial_{\sigma}  G^{2}_{\Delta=1}(X,Y) \right) + \left( \partial_{\mu} \partial_{\sigma}  G^{2}_{\Delta=1}(X,Y) \right) \left(\partial_{\nu}\partial_{\rho}  G^{2}_{\Delta=1}(X,Y) \right) \right]
    \end{aligned}
\end{equation}
where we set the $1$-point function to be zero, by definition. We can similarly compute the other terms and put them back together in \ref{eq:2ptTfinal}. We see that the central charge of the theory follows
\begin{equation}
    c\propto \sigma^4.
\end{equation}
For the free-theory, we have set $\sigma^4=1$ (recall eq \ref{eq:free-theory-limit-params}). Therefore the central charge $c$ is a fixed $O(1)$ number. Moreover, since the CFT data is expressed in terms of moments, all other moments are fixed comparing to the free scalar theory in 4d.

It takes a little bit of work to go from \ref{eq:example-term-calc} to the desired form, similar to one \ref{eq:TT_D}. As a first step, we remind ourselves that the energy-momentum tensor correlator is defined for the $D$-dimensional CFT, and the indices $\mu,\nu,\rho,\sigma$ run from $1$ to $D$. It is useful to write $G^2_{\Delta}(X,Y)$ as
\begin{equation}
    G^2_{\Delta=1}(X,Y) \sim G^2(x,y)= \frac{1}{(x-y)^2},
\end{equation}
upto a overall factor $-1/2$, which we omit, for simplicity. 
Using the following identity,
\begin{equation}\label{eq:G2deriv}
    \begin{aligned}
   \frac{\partial}{\partial x_{\mu}} \frac{\partial}{\partial y_{\rho}}G^2_{\Delta=1}(x,y) & =+2 \frac{1}{(x-y)^4}\delta_{\mu\rho} -8 \frac{(x-y)_{\mu} (x-y)_{\rho}}{(x-y)^6}\\
   &=\frac{2}{(x-y)^4}\left(\delta_{\mu\rho}-4\frac{(x-y)_{\mu} (x-y)_{\rho}}{(x-y)^2}\right)\\
   &=\frac{2}{(x-y)^4}\Tilde{I}_{\mu\rho}
\end{aligned}
\end{equation}
where we have defined
\begin{equation}
    \Tilde{I}_{\mu\rho}=\delta_{\mu\rho}-4\frac{(x-y)_{\mu} (x-y)_{\rho}}{(x-y)^2}.
\end{equation}
This gives us
\begin{equation}\label{eq:example-term-calc_a}
 \langle \partial_{\mu}\varphi(X)\partial_{\nu}\varphi(X) \partial_{\rho}\varphi(Y)\partial_{\sigma}\varphi(Y)\rangle = \frac{4}{(x-y)^8} \left( \Tilde{I}_{\mu\rho} \Tilde{I}_{\nu\sigma}+ \Tilde{I}_{\mu\sigma}\Tilde{I}_{\nu\rho} \right).
\end{equation}
One can deal with the other terms in a similar way, to get the complete expression of $\langle T_{\mu\nu}(X) T_{\rho\sigma}(Y) \rangle$.

\section{Notation Conventions}
\label{app:notation}

There are many different spaces involved in this paper, and we would like to set some conventions that we attempt to follow throughout. We use
\begin{equation}
X \in \bR^{D+1,1} \qquad \qquad x := X|_{PS} \in \bR^{D}
\end{equation}
to refer to the embedding space Minkowski $(D+2)$ coordinate and CFT $D$ coordinate, respectively, and the associated metrics are the usual $(D+2)$-Minkowski and $D$-Euclidean metrics. Fields on the embedding space and Poincar\'e section are 
\begin{equation}
\Phi(X) \qquad \qquad \phi(x) := \Phi(X|_{PS}),
\end{equation}
and the associated correlators are 
\begin{align}
G^{(n)}(X_1, \ldots, X_n) &= \bk{\Phi(X_1)\dots\Phi(X_n)}\\ 
G^{(n)}(x_1, \ldots, x_n) &:= G^{(n)}(X_1|_{PS}, \ldots, X_n|_{PS}) = \bk{\phi(x_1) \cdots \phi(x_n)}. \label{eqn:cft_corr_by_restriction}
\end{align}
respectively. That is, when we refer to fields or correlators with $X$ or $x$, they denote the fields and correlators on the full Minkowski space or their restriction to the Poincar\'e section. All expectations are computed in the neural network parameter space 
\begin{equation}
    \label{eqn:app_exp}
 \bk{\mathcal{O}} := \bE_\Theta[\mathcal{O}] = \int \mathcal{D}\Theta \, P(\Theta) \,\,  \mathcal{O},
\end{equation}
where $\Theta$ here represent all of the neural network parameters. In some cases, $\Theta$ will only be the parameters of the input layer, and we will try to be explicit about the meanings of the expectations. In normal ML applications, $P(\Theta)$ is a proper probability density function. That will sometimes be the case for us, but we will also broaden our horizons (motivated in part by usual quantum mechanics) and consider it to be a function that we integrate operators against to define expectation values according to \eqref{eqn:app_exp}.

In some cases we will also study properties of rotationally invariant theories on $\bR^{D+2}$, so that we can Wick rotate to Lorentz invariant theories on the embedding space and push down to CFTs on the Poincar\'e section. In such Euclidean $(D+2)$ theories $P(\Theta)$ is rotationally invariant, rather than Lorentz invariant, and we denote $G$ and $\Phi$ with $E$ subscripts, for Euclidean. Metric contraction via either Einstein summation or by symbolic dot product 
\begin{equation}
X \cdot Y \qquad \qquad x\cdot y
\end{equation}
are interpreted in the $(D+2)$ Euclidean or Minkowski metric (assumed to be Minkowski, unless subscript $E$ in context implies Euclidean) and $(D)$-Euclidean metrics. We may used the shorthand 
\begin{equation}
X_{12}:= X_{1}\cdot X_{2} \qquad \qquad x_{12}:= x_{1}-x_{2},
\end{equation}
where in the Minkowski $(D+2)$ case we have 
\begin{equation}
X_{12}|_{PS} = -\frac12 x_{12}^2,
\end{equation}
a standard relation between embedding space and CFT coordinates.

\bibliographystyle{JHEP}
\bibliography{main}

\end{document}